\documentclass[12pt]{article}
\usepackage{graphicx}
\usepackage{subfig}
\usepackage{amssymb, color}
\usepackage{amsmath}
\usepackage{latexsym}
\newcommand{\R}{{\mathbb R}}
\newcommand{\C}{{\mathbb C}}
\newcommand{\Z}{{\mathbb Z}}
\newcommand{\N}{{\mathbb N}}

\newcommand{\bfa}{{\bf a}}
\newcommand{\bfb}{{\bf b}}
\newcommand{\bfc}{{\bf c}}

\newcommand{\bfe}{{\bf e}}
\newcommand{\bff}{{\bf f}}

\newcommand{\bfk}{{\bf k}}

\newcommand{\bfn}{{\bf n}}

\newcommand{\bfp}{{\bf p}}

\newcommand{\bfv}{{\bf v}}
\newcommand{\bfx}{{\bf x}}
\newcommand{\bfy}{{\bf y}}

\newcommand{\bfB}{{\bf B}}

\newcommand{\bfE}{{\bf E}}

\newcommand{\bfI}{{\bf I}}

\newcommand{\bfN}{{\bf N}}

\newcommand{\bfR}{{\bf R}}
\newcommand{\bfS}{{\bf S}}

\newcommand{\eps}{{\varepsilon}}

\newcommand{\beq}{\begin{equation}}
\newcommand{\eeq}{\end{equation}}
\newcommand{\beqs}{\begin{eqnarray}}
\newcommand{\eeqs}{\end{eqnarray}}

\newcommand{\calC}{{\cal C}}

\newcommand{\calF}{{\cal F}}

\newcommand{\calH}{{\cal H}}

\newcommand{\calL}{{\cal L}}

\newcommand{\calR}{{\cal R}}
\newcommand{\calS}{{\cal S}}
\newcommand{\calT}{{\cal T}}
\newcommand{\calU}{{\cal U}}

\newcommand{\J}{J_\alpha} 
\newcommand{\Jp}{J_{\alpha+1}}
\newcommand{\Jm}{J_{\alpha-1}}
\newcommand{\pl}{{\! + \! }}

\newcommand{\FSonetimesR}{\calF_{\!\!\!\, _{S^1\times\R}}}

\newtheorem{theorem}{Theorem}[section]

\newtheorem{lemma}{Lemma}[section]

\newtheorem{definition}{Definition}[section]

\begin{document}
\begin{center}
{\Huge  
Twisted X-rays: incoming waveforms yielding discrete diffraction patterns for helical structures}
\normalsize
\vspace{0.4in}

Gero Friesecke$^*$, Richard D. James$^{**}$, and Dominik J\"ustel$^*$ \\[2mm]
$^*$Faculty of Mathematics, TU Munich, {\tt gf@ma.tum.de, juestel@ma.tum.de}  \\
$^*$Department of Aerospace Engineering and Mechanics, University of Minnesota, 
     {\tt james@umn.edu} \\
\end{center}

\vspace{0.4in}

\normalsize

\noindent {\normalsize {\bf Abstract.} 
Conventional X-ray methods use incoming plane waves and result in discrete diffraction patterns when scattered at crystals. Here we find, by a systematic method, incoming waveforms which exhibit discrete diffraction patterns when scattered at helical structures. As examples we present simulated diffraction patterns of carbon nanotubes and tobacco mosaic virus.

%Such structures are abundant in nanotechnology and molecular biology, basic examples being nanotubes and filamentous viruses.

The new incoming waveforms, which we call {\it twisted waves} due to their geometric shape, are found theoretically as closed-form solutions to Maxwell's equations. The theory of the ensuing diffraction patterns is developed in detail. A twisted analogue of the Von Laue condition is seen to hold, with the peak locations encoding the symmetry and the helix parameters, and the peak intensities indicating the electronic structure in the unit cell. 

% The theory of the ensuing diffraction patterns is developed in detail. A twisted analogue of the Von Laue condition is seen to hold: the signal of a helical structure in axial direction as a function of incoming axial and angular wavenumber is concentrated in discrete peaks governed by the reciprocal helical lattice. The unit cell electron density can then be recovered, up to a phase problem, from the peak intensities. 

If suitable twisted X-ray sources can in the future be realized experimentally, it appears from our mathematical results that they will provide a powerful tool for directly determining the detailed atomic structure of numerous biomolecules and nanostructures with helical symmetries. This would eliminate the need to crystallize those structures or their subunits.

} %end abstract

\vspace{0.2in}

\vspace{0.4in}

\tableofcontents

\vspace{0.4in}
 
%here, the superscript $T$ indicates the transpose
\section{Introduction}
This paper explores -- at the level of modelling and simulation -– the possiblity of
novel X-ray methods for the determination of the detailed atomic structure of
highly regular but not periodic molecules. The details are worked out for helical structures. These include carbon nanotubes, the necks and tails of viruses, and many of the common proteins (actin, collagen). The quest for novel methods is motivated by the fact that current X-ray methods, while hugely successful, have important shortcomings. A native helical assembly of proteins either has to be broken at the outset and the proteins crystallized, which is difficult and may lead to non-native forms; or one uses X-ray fiber diffraction, which resolves only the axial but not the angular symmetry into sharp peaks. 
\\[2mm]
Roughly, our idea is the following. Conventional X-ray methods use incoming plane waves
\begin{equation} \label{intro:PW}
      \bfE(\bfx,t)=\bfn e^{i(\bfk\cdot \bfx-\omega t)}, \;\;\; \bfB(\bfx,t)=\mbox{$\frac{1}{\omega}$}(\bfk\times \bfn) e^{i(\bfk\cdot \bfx-\omega t)},
\end{equation}
and result (in the relevant regime of X-ray wavelength $<<$ sample diameter $<<$ distance of detector from sample, Fresnel number $<<$ 1) in the outgoing field 
\begin{equation} 
   \bfE_{out}(\bfx,t) = - \frac{const}{|\bfx-\bfx_c|} \,  \bfn' e^{i(\bfk'(\bfx)\cdot \bfx-\omega t)} \int_{\Omega} e^{-i (\bfk'(\bfx)-\bfk)\cdot \bfy}\rho(\bfy)\, d\bfy,
\end{equation}
with polarization vector $\bfn'=(\bfI-\frac{\bfk'}{|\bfk'|}\otimes \frac{\bfk'}{|\bfk'|})\bfn$ and outgoing wavevector $\bfk'(\bfx)=|\bfk|\frac{\bfx-\bfx_c}{|\bfx-\bfx_c|}$. Here $\rho$ is the electron density of the illuminated sample and $\bfx_c$ denotes a typical point in the sample.

The emergence of the Fourier transform in (2), and its amazing properties regarding constructive/destructive interference, underlie the power of X-ray methods for periodic structures. One can see from (2) and its derivation that the Fourier integral kernel $e^{-i(\bfk'-\bfk)\cdot \bfy}$ is directly arising from the assumption of a plane-wave source (1). %This fact is missing in X-ray books, which start by assuming some variant of (2), and missing in electrodynamics books, which never get to (2). 
Other sources would give other kernels. This suggests the following line of research: {\it design the incoming radiation (as a solution of Maxwell's equations) 
such that the kernel interacts with highly symmetric but non-crystalline structures with the same dramatic properties of constructive/destructive interference as 
occurs in the periodic case.}

This design problem for the incoming waves can be formalized into the following mathematical problem: {\it find time-harmonic solutions to Maxwell's equations which are simultaneous eigenfunctions of a continuous extension of the generating symmetry group of the structure.} Why this is a good formalization is a long story, told in Section \ref{sec:design}. 
\\[2mm]
For discrete translation groups, which are the generating symmetries of crystals, we show that the design problem is solved precisely by the plane waves used in classical X-ray methods. This is a new characterization of plane waves. It explains why plane waves are right for crystals.

\begin{figure}[http!] 
\begin{center}
  \includegraphics[width=0.6\textwidth]{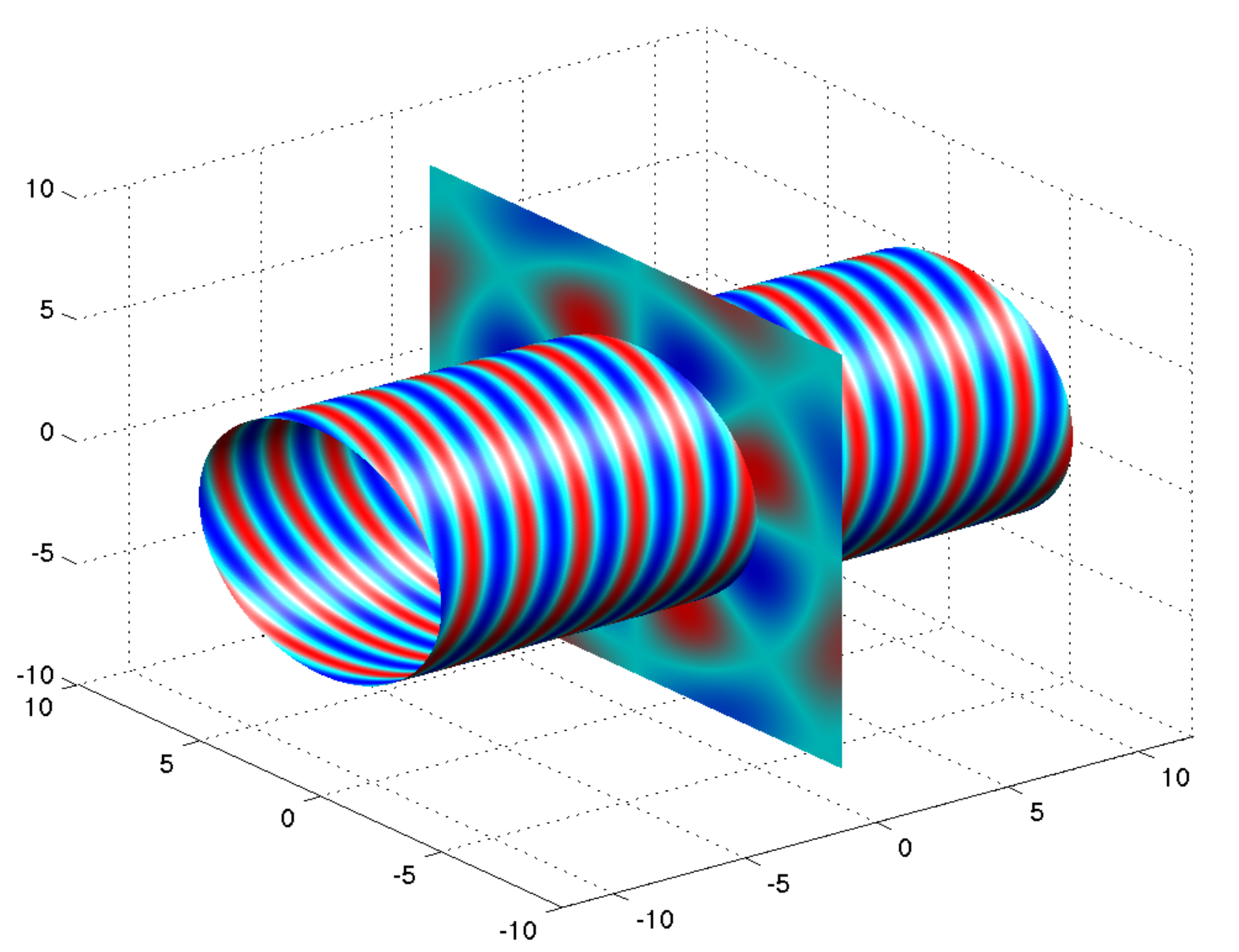}
\end{center}
\vspace*{-6mm} 

\caption{A twisted wave with angular, axial and radial wavenumber $(\alpha,\beta,\gamma)=(5,3,1)$. The plot shows the real part of the first component of the wave. Red: positive values; Blue: negative values. Note the helical shape of the level sets restricted to a co-axial cylinder, and the Bessel pattern perpendicular to the axis. 
\label{fig:TW}
} %end caption
\end{figure}

For helical symmetry groups, the design problem can also be completely solved. The ensuing family of incoming waves is 
\beq \label{intro:TW}
  \bfE(r,\varphi,z,t) = e^{i(\alpha\varphi+\beta z-\omega t)} \, 
  \begin{pmatrix} \cos\varphi & -\sin\varphi & 0 \\
                              \sin\varphi &  \cos\varphi & 0 \\
                              0          &       0     & 1 
  \end{pmatrix} \,  
  \begin{pmatrix}  
 \mbox{$\frac{n_1+in_2}{2}$} & \mbox{$\frac{n_1-in_2}{2}$} & 0 \\
 \mbox{$\frac{n_2-in_1}{2}$} & \mbox{$\frac{n_2+in_1}{2}$} & 0 \\
  0 & 0 & n_3     
     \end{pmatrix}  
  \begin{pmatrix}  \Jp(\gamma r) \\ \Jm(\gamma r) \\ \J(\gamma r)
  \end{pmatrix},
\eeq
where $(\alpha,\beta,\gamma)\in\Z\times\R\times(0,\infty)$ is a parameter vector analogous to the wavevector $\bfk$ in \eqref{intro:PW}, $\bfn\in\C^3$ is a polarization vector which must satisfy $(0,\gamma,\beta)\cdot \bfn = 0$, and the frequency $\omega$ is given by $\omega = c|(0,\gamma,\beta)|$. The cartesian vector $(0,\gamma,\beta)$ has a simple physical meaning which will emerge in Section \ref{sec:Fourier}. The $\J$ are Bessel functions, $(r,\varphi,z)$ are cylindrical coordinates with respect to the helical axis, and $\bfE_0$ is the cartesian field vector, with the third component corresponding to the axial direction. 
We call the electric fields \eqref{intro:TW} {\it twisted waves}. Figure \ref{fig:TW} shows the twisted wave with parameter vector $(\alpha,\beta,\gamma)=(5,3,1)$ and polarization vector $\bfn=(1,0,0)$. 

Thus twisted waves consist of four factors: a scalar plane wave on the cylinder;
a rotation matrix which rotates the field direction along with the base point;
a somewhat mysterious polarization tensor which depends on the polarization vector;
and a vector of three Bessel functions of neighbouring order. 

We remark that, while the axial component is just a scalar cylindrical harmonic, the twisted wave \eqref{intro:TW} is {\it not} a cylindrical vector harmonic, except for special choices of $\bfn$.\footnote{The latter fields, introduced by Hansen \cite{Hansen}, are defined as $\mbox{curl}\, (\bfa \psi)$ and $\mbox{curl}\,\mbox{curl}\, (\bfa\psi)$, where $\psi$ is a scalar cylindrical harmonic and $\bfa\in\R^3$ is a fixed ``pilot vector''. This construction provides useful basis functions for the time-harmonic Maxwell equations, but it does not capture the geometric behaviour of the polarization of twisted waves.} 
\\[2mm]
The helical shape of the level sets in Figure \ref{fig:TW} suggests that for suitable values of the angular and axial wavenumber, twisted waves can induce resonant electronic oscillations of every single molecule in a structure with helical architecture. Thus one can hope for diffraction intensities which strongly depend on the twisted wave parameters. 

At least in axial direction, the radiation scattered by a helical structure indeed exhibits sharp discrete peaks with respect to the radiation parameters $\alpha$ and $\beta$. More precisely: {\it the signal of a helical structure in axial direction vanishes unless the angular/axial wavenumbers of the twisted wave minus the axial wavenumber of the outgoing wave belong to the reciprocal helical lattice shifted left or right by precisely one angular wavenumber.} This is an analogue of the Von Laue condition, but waves and structure are curved. The shifts come from the fact that the polarization direction of a twisted wave rotates along with the base point. Details are given in Section \ref{sec:TVL}. Moreover, as in X-ray crystallography, the unit cell electron density can be recovered, up to a scalar phase problem, from the peak intensities. A further attractive feature is that the outgoing signal is invariant under axial translations and rotations of the structure.\footnote{By comparison, the signal poduced by fiber diffraction as described by the Cochran-Crick-Vand formula \cite{CCV52} is not invariant under axial rotations, causing well-known difficulties in the interpretation of fiber diffraction images.}

\begin{figure}[http!] 
\begin{center}
   \includegraphics[width=0.9\textwidth]{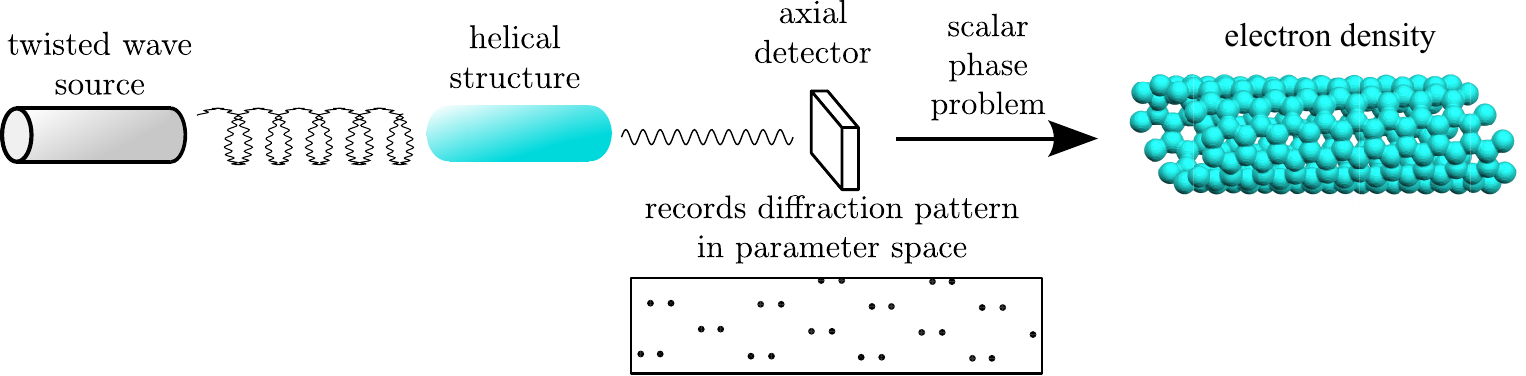}
\end{center}
\vspace*{-3mm}

\caption{Hypothetical set-up of structure analysis with twisted X-rays.}
\label{fig:setup}
\end{figure}

These results suggest a hypothetical set-up of structure analysis with twisted X-rays. 
Send a twisted wave towards a co-axial helical structure. Use a detector further along the axis to record the diffracted intensities as a function of the incoming radiation parameters. Solve a scalar phase problem to infer the electron density. See Figure 
\ref{fig:setup}. 

%\textcolor{red}{- Advantages compared to classical X-ray and fiber diffraction (...sharp peaks only in half of the symmetry-related DOF's; we get all)}

Twisted X-ray waves, \eqref{intro:TW}, are a theoretical proposal. Intriguingly, similar waveforms have been experimentally realized, such as optical higher-order Bessel beams \cite{AD00} and photons and beams carrying an angular wave factor $e^{i\alpha\varphi}$ \cite{AB92,HD92,MTT07}. Clarifying the precise relationship to twisted waves would require an understanding of the full electromagnetic field of the produced modes. A related recent development is the creation of electron vortex beams \cite{UT10,VTS10}.    
%
%
%
% Allen et al. 1992 say they have produced a `Laguerre beam'
% Heckenberg et al. 1992 say they have produced `helical waves'
% Both of them try to interpret their waves via scalar models only

The plan of the paper is as follows. Sections 2--4 extend the standard scalar Fourier transform model for plane-wave diffraction to a vector-valued electromagnetic model needed to treat general incoming waveforms. Sections 6--9 formulate the radiation design problem, and derive plane waves and twisted waves from it. Sections 10--12 develop the theory of diffraction patterns of twisted X-rays. Finally, in Section 13 we present simulated diffraction patterns of a carbon nanotube and of tobacco mosaic virus.

\section{The time-harmonic Maxwell equations}
Incoming electromagnetic waves will be sought as solutions to Maxwell's equations in vacuum, 
\beq \label{M}
   \frac{1}{c^2}\frac{\partial \bfE}{\partial t} = {\rm curl}\, \bfB, \;\; {\rm div}\, \bfE=0, \;\;  
   \frac{\partial\bfB}{\partial t} = - {\rm curl}\, \bfE,  \;\; {\rm div}\, \bfB = 0,
\eeq
which are time-harmonic, that is to say 
\beq \label{TH}
   \bfE(\bfx,t) = \bfE_0(\bfx)e^{-i\omega t}, \;\; \bfB(\bfx,t) = \bfB_0(\bfx)e^{-i\omega t}
\eeq
for some $\omega>0$. Here $\bfE$ and $\bfB$ are the electric and magnetic fields and $c$ is the speed of light.
SI units are used throughout. The fields are defined on $\R^3\times[0,\infty)$ and take values in $\C^3$.   
\\[2mm]
The ansatz \eqref{TH} reduces Maxwell's equations \eqref{M} to 
\begin{eqnarray}
   && \Delta \bfE_0 = - \frac{\omega^2}{c^2} \bfE_0, \;\; {\rm div}\, \bfE_0 = 0, \label{E0} \\
   && \bfB_0 = -\frac{i}{\omega} {\rm curl}\, \bfE_0. \label{B0}
\end{eqnarray}
We are interested in bounded solutions to \eqref{E0}. We note that bounded weak solutions, 
that is to say vector fields in the function space $L^\infty(\R^3;\C^3)$ which solve \eqref{E0} 
in the sense of distributions, are automatically bounded infinitely differentiable vector fields which 
satisfy \eqref{E0} classically. 
\\[2mm]
A basic example of time-harmonic radiation, and the one used in classical X-ray crystallography, are plane
waves
\beq \label{PW}
    \bfE_0(\bfx) = \bfn e^{i\bfk_0\cdot \bfx}, \;\; \bfB_0(\bfx) = \frac{1}{\omega} (\bfk_0\times \bfn)e^{i\bfk_0\cdot \bfx}
\eeq 
which solve \eqref{E0}--\eqref{B0} if the wavevector $\bfk_0\in\R^3$ and the polarization vector $\bfn\in\C^3$
satisfy
\beq \label{PWeqs}
    |\bfk_0|=\frac{\omega}{c}, \;\; \bfk_0\cdot\bfn = 0.
\eeq
Note that by the first of these relations, the time frequency $\omega$ in \eqref{TH} fixes the spatial wavelength
$\lambda=2\pi/|\bfk_0|$ of plane-wave radiation as
\beq \label{wavelength}
    \lambda = 2\pi\frac{c}{\omega}.
\eeq 

The more general form \eqref{TH} is important when scattering from non-crystalline structures is under 
consideration. In fact the notion of wavelength and the relation \eqref{wavelength} remain meaningful 
in this case. This is because the Fourier transform 
$
    \hat{\bfE}_0(\bfk)=\int_{\R^3}e^{-i\bfk\cdot\bfx}\bfE_0(\bfx)d\bfx
$ 
is, due to eq. \eqref{E0}, 
supported on the wavevector sphere $|\bfk|=\omega/c$, and hence any solution $\bfE_0$ can be thought of as a superposition 
of plane waves with wave vectors of equal magnitude $\omega/c$.

In X-ray crystallography, due to the goal of atomic resolution,
only $\lambda$'s comparable to interatomic distances, i.e. below a few Angstrom, are useful. Electromagnetic 
waves in this regime are known as {\it hard X-rays}. 

\section{Electromagnetic model for the diffracted radiation} \label{model}

The standard scalar Fourier transform model (or {\it oscillator model}) of X-ray diffraction patterns which underlies the crystallography and biocrystallography literature is insufficient for our purposes. This is because it neglects the polarization of the incoming wave from the outset. However, fluctuating polarization directions turn out to be a necessary feature of any incoming field which can induce resonant electronic oscillations in a helical (or any other symmetric but non-crystalline) structure. As a consequence we need a full electromagnetic (vector-valued) extension of the standard model. 

Before presenting this extension, we briefly recall the standard scalar model, for comparison and background.  
\\[2mm]
{\bf Oscillator model.} As described in many texts (see e.g. \cite{Ashcroft}), 
one starts from a simplified picture which involves only scalar waves instead of electromagnetic fields and bypasses Maxwell's equations. A scalar incoming plane wave $e^{i(\bfk_0\cdot \bfx-\omega t)}$ hitting an electron at position 
$\bfy$ is assumed to induce a circular wave with same wavelength and same phase at the point $\bfy$, 
\beq\label{osc-single}
   const \, e^{i(|\bfk_0| \, |\bfx-\bfy|-\omega t)}e^{i\bfk_0\cdot \bfy}.
\eeq
Suppose that the incoming wave travels at the speed of light, $|\bfk_0|=\omega/c$; let $\bfx_c$ be a typical point in the sample; approximate the phase $|\bfk_0| |\bfx-\bfy|$ of the outgoing circular wave at a point $\bfx$ in the far field by
$\bfk'(\bfx)\cdot(\bfx-\bfy)$, with direction-dependent outgoing wavevector
\beq \label{k'}
   \bfk'(\bfx) = \frac{\omega}{c} \frac{\bfx-\bfx_c}{|\bfx-\bfx_c|}; 
\eeq
and take a superposition of electrons with density $\rho(\bfy)$. This yields the scalar outgoing field
\beq \label{osc}
  \psi_{out}(\bfx,t)
  = const \, e^{i (\bfk'(\bfx)\cdot \bfx - \omega t)} \int_{\R^3} \rho(\bfy) e^{-i(\bfk'(\bfx)-\bfk_0)\cdot\bfy} d\bfy.
\eeq
Note that the integral appearing above, 
\beq \label{str}
   f(\bfk'-\bfk_0) = \int_{\R^3} e^{-i(\bfk'-\bfk_0)\cdot \bfy} \rho(\bfy) \, d\bfy = \hat{\rho}(\bfk'-\bfk_0),
\eeq 
is just the Fourier transform of the electron density, evaluated at the difference 
$\bfk'-\bfk$ of outgoing and incoming wavevector. This integral encodes the density of the structure 
one wants to image, and is known as the {\it structure factor} or {\it form factor} or {\it scattering factor}. The squared amplitude $I(\bfx)=|\psi(\bfx,t)|^2$ of the scalar field \eqref{osc} is then taken as a model for the measured intensity of the outgoing electromagnetic radiation. From \eqref{osc}, one obtains
\beq \label{I-osc}
     I(\bfx;\bfk_0)=|\psi_{out}(\bfx,t)|^2=|\hat{\rho}(\bfk'(\bfx)-\bfk_0)|^2.
\eeq
Thus the ``intensity of diffracted radiation'' at the detector position $\bfx$ is the 
absolute value squared of the Fourier transform of $\hat{\rho}$, evaluated at the difference
$\bfk'(\bfx)-\bfk_0$ between outgoing and incoming wavevector. To emphasize the dependence of $I$ on both 
the observation point (which determines the outgoing wavevector) and the parameters of the incoming
wave we have used the notation $I(\bfx;\bfk_0)$. 

In particular, the data gathered from a
sufficiently large set of observation points $\bfx$ and incoming wave parameters 
$\bfk_0$ delivers the abstract data set 
\beq \label{iman}
   I(\bfk) = |\hat{\rho}(\bfk)|^2, \;\; \bfk\in\R^d, 
\eeq
and the X-ray interpretation problem, i.e. the task of inferring atomic structure, $\rho$, from X-ray diffraction data, $I$, reduces to the phase problem for the Fourier transform. Note that the data set \eqref{iman} is a function on the dual (wavevector) space of the physical space on which the density is defined. The abstract ``diffraction intensity'' or ``diffraction spectrum'' or ``diffraction measure'' \eqref{iman} constitutes the starting point of previous mathematical work on X-ray diffraction of crystals \cite{Strichartz,Friesecke} and quasicrystals \cite{Hof, BaakeMoody, BaakeGrimm, BaakeGrimmBook}.

The semi-empirical oscillator model \eqref{k'}, \eqref{osc}, \eqref{I-osc} is sufficient for a great deal of X-ray physics. In particular, it accounts for the fluctuating phases of incoming plane-wave radiation \eqref{PW}.
\\[2mm]
{\bf Electromagnetic model.} 
A proper vector-valued expression for the outgoing electric field which improves 
the model \eqref{osc} and can deal with fluctuating incoming polarization directions can be obtained as follows. Start from the well known Lienard-Wiechert fields (see e.g. \cite{Griffiths}) of a moving point charge driven by any given incoming field; pass to a non-relativistic (weak-field) limit; take a superposition of charges; and make a far field approximation corresponding to the regime X-ray wavelength $<<$ sample diameter $<<$ distance of observation point from sample, Fresnel number $<< 1$.
The details can be found in our companion paper \cite{FJJ2} and we only give the resulting expression for the diffracted electromagnetic field:
%\footnote{Even in the special case of the incoming field being a standard plane wave and the density consisting of a single electron, $\rho=\delta_0$, we could not find this result in X-ray textbooks. Expressions for the outgoing electromagnetic field were either missing, incomplete, or incorrect, even in the best texts, with the mistakes differing from book to book (for examples see \cite{FJJ2}).} 
\beq \label{out}
  \begin{array}{lll} 
   \bfE_{out}(\bfx,t)
     & =   & - \; c_{e\ell} \, \frac{e^{i( \bfk'(\bfx) \cdot \bfx - \omega t )}} {|\bfx-\bfx_c|}
      \left(\bfI - \frac{\bfk'(\bfx)}{|\bfk'(\bfx)|} \otimes  \frac{\bfk'(\bfx)}{|\bfk'(\bfx)|}\right) 
         \int_{\R^3} \bfE_0(\bfy) \rho(\bfy) \, e^{-i\bfk'(\bfx)\cdot \bfy} d\bfy, \\[1mm]
  \bfB_{out}(\bfx,t)
     & =  & \frac{1}{\omega} \, \bfk'(\bfx) \, \times \, \bfE_{out}(\bfx,t), 
  \end{array}
\eeq
with direction-dependent outgoing wavevector familiar from the oscillator model, 
\beq \label{k''}
    \bfk'(\bfx) = \frac{\omega}{c} \frac{\bfx-\bfx_c}{|\bfx-\bfx_c|}. 
\eeq
Here $\bfE_0(y)e^{-i\omega t}$ is the incoming electric field, a solution to the time-harmonic Maxwell equation 
\eqref{E0}, $\rho\, : \, \R^3\to\R$ is the electron density of the sample,
$\bfx_c$ is a typical point in the sample, and $c_{e\ell}$ is a universal constant depending, 
among other things, on the charge and mass of the electron. Moreover $\bfI$ is the 3$\times$3 identity
matrix and $\bfa\otimes\bfb$ denotes the 3$\times$3 matrix $A$ with entries $A_{ij} = a_i b_j$, where $\bfa$
and $\bfb$ are any two vectors in $\R^3$. 

If the incoming electromagnetic field is replaced by its real part, as it properly should to model physical
incoming X-rays, the diffracted radiation is given by the real part of \eqref{out}.  

According to the model \eqref{out}--\eqref{k''}, the outgoing wavevectors $\bfk'$ have the same length as the incoming 
wavevectors $\bfk$ (recall from the previous section that any solution $\bfE_0$ to eq. \eqref{E0} can be 
viewed as a superposition of plane waves with wavevectors $\bfk$ of equal length $|\bfk|=\omega/c$), 
\beq \label{cons}
   |\bfk'| = |\bfk|.
\eeq
This relation has the important microscopic physical interpretation that the photon energy is conserved in the 
scattering. Note that the energy of a photon with wavevector $\bfk$ is $E=|\bfp| c$, where $\bfp=\hbar\bfk$ is
the photon momentum. Thus the model \eqref{out} corresponds to {\it elastic} or {\it Thomson} scattering. 

In the special case of incoming plane-wave radiation \eqref{PW}--\eqref{PWeqs}, eq. \eqref{out} for the
electromagnetic field reduces to
\beq \label{PWout}
   \bfE_{out}(\bfx,t)
   =  - \, c_{e\ell} \, \frac{e^{i( \bfk'(\bfx) \cdot \bfx - \omega t )}} {|\bfx-\bfx_c|} \bfn'(\bfx) \, 
   \int_{\R^3} \rho(\bfy) \, e^{-i(\bfk'(\bfx)-\bfk_0)\cdot \bfy} d\bfy, \;\;  
   \bfB_{out}(\bfx,t) = \frac{1}{\omega}\bfk'(\bfx) \times \bfE_{out}(\bfx,t),
\eeq
with outgoing direction-dependent polarization vector
\beq \label{n'}
   \bfn'(\bfx) = \left(\bfI - \frac{\bfk'(\bfx)}{|\bfk'(\bfx)|} \otimes  \frac{\bfk'(\bfx)}{|\bfk'(\bfx)|}\right)\bfn.
\eeq

In X-ray experiments the electromagnetic fields $\bfE$ and $\bfB$ cannot be measured directly; a detector
only records the scalar field intensity. The latter is defined (see e.g. \cite{Griffiths}) 
as the {\it time average of the absolute value of the Poynting vector}, and has the physical 
dimension of power transferred per unit area, i.e. energy transferred per unit area per unit time. In formulae,
\beq \label{I}
       I(\bfx) = \lim_{T\to\infty} \frac{1}{T}\int_0^T |\bfS(\bfx,t)| \, dt, \;\;\; \bfS(\bfx,t) = \frac{1}{\mu_0}\,
       \bfE(\bfx,t)\times\bfB(\bfx,t).
\eeq
Here and below, $\mu_0$ and $\eps_0$ are the magnetic and electric constants, which are related to the speed of light by the equation $\eps_0\mu_0=1/c^2$, and $\bfS$ denotes the Poynting vector. Note that the latter is the flux vector for the energy density $e$ in Maxwell's equation, that is to say real-valued solutions to \eqref{M} satisfy
\beq \label{en}
   \frac{\partial}{\partial t}e + \mbox{div}\, \bfS = 0, \;\; \mbox{where} \; e(\bfx,t)=\frac12
   \left(\eps_0|\bfE|^2 + \frac{1}{\mu_0} |\bfB|^2\right).
\eeq
Here the restriction to physical, real-valued solutions is essential, and we do not recommend that the above 
definition of the Poynting vector be applied to complex fields. In Appendix 1 we evaluate expression \eqref{I}
for the physical diffracted radiation, given by the real part of the fields $\bfE_{out}$ and
$\bfB_{out}$ in eq. \eqref{out}. The result, i.e. the 
intensity of the real part of the elecromagnetic field \eqref{out} at the observation point $\bfx$, is
\beq \label{Iout}
        I(\bfx) = \frac{c\,\eps_0}{2} \, |\bfE_{out}(\bfx,t)|^2,
\eeq
where $\bfE_{out}$ is the complex electric field in \eqref{out} (note that its absolute value is independent of $t$).
The at first sight mysterious appearence of the {\it complex} field amplitude comes from the time averaging in 
\eqref{I}, as is clear from the derivation in Appendix 1.
\\[2mm]
In the plane-wave case, the intensity can be read off from eqs. \eqref{PWout} and \eqref{Iout}. Noting that $\bfn'$ is the projection of $\bfn$ onto the orthogonal subspace 
of $\bfk'$, and assuming for the moment that $\bfn$ is real up to a phase factor (i.e., the wave is linearly polarized), we have
\beq \label{absn'}
   |\bfn'| = \sin\left(\angle(\bfn,\bfk')\right)\, |\bfn|,
\eeq
and hence
\beq \label{IPWout}
   I(\bfx;\bfk_0) = \frac{c \, \eps_0 \, c_{e\ell}^2}{2} \, \frac{1}{|\bfx-\bfx_c|^2} \, 
                    \sin^2\left(\angle(\bfn,\bfk'(\bfx))\right) \, |\bfn|^2
                    |\hat{\rho}(\bfk'(\bfx)-\bfk_0)|^2. 
\eeq
(Here we have used the same notation as in \eqref{I-osc} to emphasize the dependence on the incoming wavevector $\bfk_0$.) If $\bfn$ is complex, the $\sin^2$ factor has to be replaced by $|\bfn'(\bfx)|^2$, with $\bfn'$ given by \eqref{n'}.   
In particular, the electromagnetic model \eqref{out}--\eqref{k''} specialized to plane waves
recovers in a natural way the well known fact -- missed by the oscillator model -- that scattering along the polarization direction of the incoming X-ray beam, corresponding to $\bfk'(\bfx)$ being parallel to $\bfn$, 
is completely suppressed. In the special case of a single electron, $\rho=\delta_0$, expression 
\eqref{IPWout} agrees with what is typically assumed as a starting point in X-ray books 
(see e.g. \cite{Nielsen} Section 1.2). Note also that the last factor in \eqref{IPWout} 
is precisely the intensity of the semi-empirical oscillator model, i.e. 
the absolute value squared of the ``structure factor'' \eqref{str}. 

Although the arguments in \cite{FJJ2} leading to \eqref{PWout}--\eqref{n'} are purely classical, the result agrees
(up to the decay factor $1/|\bfx-\bfx_c|$, and granting the identification of macroscopic outgoing directions
with microscopic photon wavevectors) with the result derived in the beautiful tutorial article \cite{Santra}
from perturbative nonrelativistic quantum electrodynamics (NRQED). In particular, NRQED confirms equations 
\eqref{k''} and \eqref{n'}. 

In the sequel, we shall work with the electromagnetic model \eqref{out}--\eqref{k''}. 
\section{Generalized structure factor}
From the point of view of contemplating novel forms of incoming radiation, 
the most important outcome from the electromagnetic model 
\eqref{out}--\eqref{k''} is a prescription for how the ``structure factor'' \eqref{str} 
needs to be modified when the incoming X-rays are not plane waves. 
The corresponding term in \eqref{out} is the integral term 
\beq \label{genstr}
    \bff_{\bfE_0}(\bfk') = \int_{\R^3} \bfE_0(\bfy)\, \rho(\bfy) \, e^{-i\bfk'\cdot \bfy} d\bfy. 
\eeq
We call this term the {\it generalized structure factor} with respect to the incoming electric field $\bfE_0$. The latter may be any bounded solution to the time-harmonic Maxwell equation \eqref{E0}. Expression \eqref{genstr} is a vector-valued generalization of the familiar structure factor 
\begin{equation} \label{oldstr}
              f(\bfk'-\bfk_0) = \int_{\R^3} \rho(\bfy) \, e^{-i(\bfk'-\bfk_0)\cdot \bfy} d\bfy 
              \;\;\; (\mbox{eq. \eqref{str}, Section \ref{model}})
\end{equation}
for classical (plane-wave) diffraction, and determines the amount of constructive or destructive interference of the outgoing waves in the direction $\bfk'$. Note that the interference in \eqref{genstr} depends not just on the incoming and outgoing phases, but also on the incoming polarization directions.

In the plane-wave case \eqref{PW}, that is to say $\bfE_0(\bfy)=\bfn e^{i\bfk_0\cdot \bfy}$, the exponential factor 
from the incoming wave nicely combines with the exponential factor from \eqref{genstr}, 
and the generalized structure factor reduces to just the classical structure factor \eqref{str}, multiplied by the constant incoming polarization vector: 
\beq \label{genstrPW}
   \bff_{\bfE_0}(\bfk') = \bfn \, \int_{\R^3} \rho(\bfy) \, e^{-i(\bfk'-\bfk_0)\cdot\bfy} d\bfy
                              = \bfn \, f(\bfk'-\bfk_0).
\eeq

It will be useful to interpret \eqref{genstr} mathematically as an integral transform that depends on the 
incoming field $\bfE_0$ and maps the electronic charge density $\rho$ to a vector field on the dual (wavevector) space. It generalizes the Fourier transform which arises in the oscillator model, and might be called the {\it radiation transform} of $\rho$ with respect to $\bfE_0$. This transform is defined by    
\beq \label{rad}
  (\calR_{\bfE_0}\rho)(\bfk) = \int_{\R^3} \bfE_0(\bfy)\rho(\bfy)e^{-i\bfk \cdot \bfy} d\bfy \;\;\;(\bfk\in\R^3).
\eeq
Here, as before, $\bfE_0$ may be any bounded solution to the time-harmonic Maxwell equation \eqref{E0}. 
Eq. \eqref{rad} naturally establishes $\calR_{\bfE_0}$ as a linear map from $L^1(\R^3)$ to the space 
of bounded continuous vector fields on the dual (wavevector) space $\R^3$. 
Like the Fourier transform, it can be extended to tempered distributions, 
and maps these to vector-valued tempered distributions. This extension will be useful when discussing 
diffraction in the idealized but important case of infinite helices, nanotubes and crystals.  
\section{Classical Von Laue condition}
Given a complete mathematical model of the outgoing electromagnetic field such as \eqref{out}, we can cast the Von Laue condition of classical X-ray crystallography, first discovered by Friedrich, Knipping, and Von Laue \cite{FKL12}, in the form of a mathematical theorem. 
Let $\calL$ be a Bravais lattice in $\R^3$, i.e. a set of the form 
\beq \label{Bravais}
   \calL = A \, \Z^3 \;\; \mbox{ for some invertible $3\times 3$ matrix $A$}. 
\eeq
Let $\rho$ be any smooth $\calL$-periodic function on 
$\R^3$, that is to say 
$$
   \rho(\bfx+\bfa) = \rho(\bfx) \mbox{ for all }\bfx\in\R^3, \; \bfa\in\calL.
$$
By choosing a suitable partition of unity, any such $\rho$ can be written in the form 
\beq \label{perdens}
   \rho(\bfx) = \sum_{\bfa\in\calL} \varphi(\bfx-\bfa) 
\eeq
for some smooth, rapidly decaying function $\varphi$ belonging to the Schwartz space $\calS(\R^3)$. 
%It is also of physical and mathematical interest to consider the case  $\varphi(\bfx)=\delta(\bfx - \bfx_0}$, which leads to the simplified model
%\beq \label{deltadens}
%   \rho(\bfx) = \sum_{\bfa\in\calL} \delta (\bfx - \bfx_0 - \bfa). 
%\eeq
In the sequel, we will use the following notation for delta functions common in the mathematical literature: if $\bfx_0$ is a point in $\R^3$, and $\calS$ a countable set of points in $\R^3$, we write
\beq \label{delnot}
   \delta_{\bfx_0}(\bfx) = \delta (\bfx-\bfx_0 ), \;\;\; 
   \delta_{\calS}(\bfx) =  \sum_{\bfa\in\calS} \delta(\bfx-\bfa).
\eeq
In the Von Laue condition, the reciprocal lattice of $\calL$ will naturally emerge. The latter is defined
by 
\beq \label{L'} 
   \calL' = \{\bfk\in\R^3\, : \, \bfk\cdot\bfa \in 2\pi\Z\mbox{ for all }\bfa\in\calL\},
\eeq
and is given in the case of \eqref{Bravais} explicitly by
\beq \label{L'explicit} 
    \calL' = 2\pi A^{-T} \Z^3,
\eeq
where $A^{-T}$ denotes the transpose of the inverse of the matrix $A$. The Von Laue condition can now be stated as follows. 

For simplicity, we only consider the square root of the intensity, i.e., up to trivial 
constants, the absolute value of the electric field \eqref{PWout}. This is convenient because the 
field itself, unlike its square, is a well-defined mathematical object 
for infinitely extended systems such as \eqref{perdens} or $\rho=\delta_{\calL}$, by interpreting it as a distributional Fourier transform. The powerful machinery of Fourier analysis then allows to mathematically understand in a quick way the phenomenon of discrete diffraction patterns.
\begin{theorem} {\bf (Von Laue condition).} Let $\calL$ be the Bravais lattice \eqref{Bravais}, and 
let $\rho\, : \, \R^3\to\R$ be any $\calL$-periodic density, represented in the form \eqref{perdens}. Assume that the incoming electric field is a plane wave, $\bfE_0(\bfy)=\bfn\, e^{i\bfk_0\cdot\bfy}$, 
with polarization vector $\bfn\in\C^3$ and wavevector $\bfk_0\in\R^3$ satisfying \eqref{PWeqs}, and
the outgoing radiation is given by the electromagnetic model \eqref{out}--\eqref{k''}, with corresponding 
intensity \eqref{IPWout}. Assume moreover without loss of generality that $\bfx_c=0$. Then 
\begin{eqnarray} \label{Laue}
  (I(\bfx;\bfk_0))^{1/2} &=& c_0 \, \frac{(2\pi)^3}{|\det A|} \, \frac{1}{|\bfx|} \, 
                         \Big| \bfn - 
      (\mbox{$\frac{\bfk'(x)}{|\bfk'(x)|}$}\cdot\bfn) 
       \mbox{$\frac{\bfk'(x)}{|\bfk'(x)|}$}
                         \Big| \, 
                         \left|\hat{\varphi}\left(\bfk'(\bfx) - \bfk_0\right)\right| \, 
                         \delta_{\calL'} \left( \bfk'(\bfx) - \bfk_0\right) \\
                           &=& c_0 \, \frac{(2\pi)^3}{|\det A|} \, \frac{1}{|\bfx|} \, 
                         \Big| \bfn - 
      (\mbox{$\frac{\bfk'(x)}{|\bfk'(x)|}$}\cdot\bfn) 
       \mbox{$\frac{\bfk'(x)}{|\bfk'(x)|}$}
                         \Big| \, 
                         \sum_{\bfa'\in\calL'} |\hat{\varphi}(\bfa')| \delta_{\bfa'}
                         \left(\bfk'(\bfx) - \bfk_0\right),
\end{eqnarray}
where $\bfk'(\bfx)=\frac{\omega}{c}\frac{\bfx}{|\bfx|}$ and $c_0=(\frac{c\eps_0}{2})^{1/2}c_{e\ell}$. In particular, the outgoing signal is zero unless the difference between outgoing and incoming wavevector, $\bfk'(\bfx) - \bfk_0$, 
is a reciprocal lattice vector. 
\end{theorem}
For the simple diffraction model \eqref{iman}, 
analogous results for $I^{1/2}$ were presented 
in \cite{Strichartz, Friesecke}, and a rigorous treatment of the intensity $I$ renormalized by volume was given in \cite{Hof}. 
We remark that the right hand side of \eqref{Laue} makes rigorous sense
(as a locally bounded measure) if considered as a function of the incoming wavevector $\bfk_0$ at fixed 
observation point $\bfx$, as implicitly assumed in the simple model \eqref{iman}, but not if considered 
as a function of $\bfx$ at fixed $\bfk_0$.
\\[2mm]
{\bf Proof} The central term in the outgoing field \eqref{PWout} is the structure factor $\bff_{\bfE_0}(\bfk') = 
\bfn \hat{\rho}(\bfk'-\bfk_0)$, which is well-defined as a distributional Fourier transform. It is convenient
to re-write formula \eqref{perdens} as the convolution 
$$
    \rho = \varphi * \delta_{\calL},
$$ 
where $(f*g)(\bfx)=\int_{\R^3}f(\bfx-\bfy)g(\bfy)\, d\bfy$. By the generalized Poisson summation formula
(see e.g. \cite{Friesecke}), the Fourier transform of $\delta_{\calL}$ is
\beq\label{Poisson}
    \widehat{\delta_{\calL}} = \frac{(2\pi)^3}{|\det A|} \delta_{\calL'}.
\eeq
The Fourier calculus rule $\widehat{f*g}=\hat{f}\, \hat{g}$ now gives
$$
  \hat{\rho} = \frac{(2\pi)^3}{|\det A|} \hat{\varphi} \, \delta_{\calL'}, \hspace*{1cm} 
  f(\bfk'(\bfx)-\bfk_0) = \frac{(2\pi)^3}{|\det A|} \, (\hat{\varphi}\delta_{\calL'})
   \left(\bfk'(\bfx) - \bfk_0\right).
$$
Since the expression for the structure factor is, by inspection, a locally bounded measure, its absolute value is well defined (see e.g. \cite{Di75}), and corresponds in the above case to replacing the factor $\hat{\varphi}$ by its absolute value. The result now follows immediately from \eqref{IPWout}. 
\section{The euclidean group of isometries}
A key role in the following is played by the euclidean group of isometries
$$
    E(3) = \{ (\bfR|\bfc) \, : \, \bfR\in O(3), \, \bfc\in\R^3 \},
$$
where $O(3)=\{\bfR\in M^{3\times 3} \, : \, \bfR^T\bfR=\bfI\}$ is the orthogonal group. Elements of $E(3)$ 
act on points in three-dimensional euclidean space $\R^3$ as 
$$
    (\bfR|\bfc)\bfx = \bfR\bfx + \bfc.
$$
The natural induced action on vector fields $\bfE\, : \, \R^3\to\C^3$ is
\beq \label{action}
    \left( (\bfR|\bfc)\bfE\right) (\bfx) = \bfR\bfE\left(\bfR^{-1}(\bfx-\bfc)\right).
\eeq
Note that $\bfR^{-1}(\cdot-c)$ is the inverse element $(\bfR|\bfc)^{-1}$. The action \eqref{action} preserves solutions to Maxwell's equations. 
\\[2mm]
Important continuous subgroups are 
\beq \label{trans}
   \calT = \{ (\bfI|\bfc) \, : \, \bfc\in\R^3\}, \;\; \bfx\mapsto\bfx+\bfc \;\; \mbox{(translation group)}
\eeq
and
\beq \label{hel}
   \calH_{\bfe} = \{ (\bfR|\tau\bfe) \, : \, \bfR\in SO(3), \, \bfR\bfe=\bfe, \, \tau\in\R\}, \;\;
                   \bfx \mapsto \bfR\bfx + \tau\bfe \;\;\mbox{(helical group with axis $\bfe$)},
\eeq
where $\bfe$ is a given unit vector in $\R^3$. Here $SO(3)$ denotes the special orthogonal group, i.e. the 
elements of $O(3)$ with determinant $1$. As we shall see shortly, the former subgroup is associated with classical
plane-wave X-ray methods for crystals, and the latter with twisted X-rays. 
\section{The design equations} \label{sec:design}
We now have a second look at plane waves. Why are they the right radiation to use for the 
analysis of crystals?

Every researcher interested in X-ray diffraction is familiar with the Von Laue 
condition: the outgoing signal of a plane wave scattered at a crystal concentrates on a discrete set of outgoing directions, while in between it is suppressed by destructive interference. 
But suppose plane waves were not given to us a priori, as the radiation emitted by conventional X-ray tubes.
Would we know how to come up with them if our goal was to achieve a nice diffraction pattern? 
What, exactly, is the ``connection'' between a particular family of solutions to Maxwell's equations on the one hand, and point sets with crystalline order on the other?

The connection is that plane waves and crystals have {\it matching symmetries}. 
By this we do not mean that they have the same symmetries. Plane waves have a larger, ``continuous'' family of symmetries. We first explain 
% describe
this informally,
then make it precise in group-theoretical language, then generalize beyond crystals. 
\\[2mm]
Start with a crystal, i.e. a structure with atomic positions
\beq \label{pos1}
   \calS = \{\bfx_0^{(\nu)} + \bfa \, : \, \bfa\in\calL, \; \nu=1,..,M\}, \;\; 
   \calL = \{i \bfv_1 + j \bfv_2 + k \bfv_3\, : \, i,j,k\in\Z\} = A\,\Z^3,
\eeq
where $\bfx_0^{(1)},..,\bfx_0^{(m)}$ are the positions of finitely many reference atoms, 
$\bfv_1$, $\bfv_2$, $\bfv_3$ are linearly independent vectors in $\R^3$, and $A$ is the matrix with columns given by these vectors.
Thus the atom positions are obtained by translating finitely many reference atoms by each element of the lattice  $\calL$. 
Mathematically, this means that the crystal is the ``orbit'' of a finite set of points under a discrete group $\calL$ of translations, and in particular that each element of $\calL$ is a symmetry of the crystal, i.e. maps it to itself. The discrete translation group $\calL$ is the {\it generating symmetry group} of the crystal. 

We now look at plane waves, 
\beq \label{PW0}
   \bfE_0(\bfy) = \bfn e^{i\bfk_0\cdot \bfy}, \;\; \bfn\in\C^3, \;\; \bfk_0\in\R^3, \;\; 
                  \bfk_0\cdot\bfn = 0.
\eeq
Plane waves also have a translation symmetry. Their values at different positions just differ by phase factors, 
\beq \label{DE0}
   \bfE_0(\bfx_0 + \bfa) = \mbox{(phase factor depending on $\bfa$)} \, \bfE_0(\bfx_0)
   \; \mbox{for all }\bfa\in\R^3,
\eeq
with
\beq \label{phase}
    \mbox{(phase factor in \eqref{DE0})} = e^{i\bfk_0\cdot\bfa}.
\eeq
Mathematically, eq. \eqref{DE0} means that the wave, a vector field on $\R^3$, is an ``eigenfunction'' 
% under the operation of translating
of the operator which translates 
a vector field by a vector $\bfa\in\R^3$, 
%$\bfE_0\mapsto\bfE_0(\cdot - \bfa)$. These operations
$T_{-\bfa} \, : \, \bfE_0\mapsto\bfE_0(\cdot + \bfa)$. These operators 
form a continuous group which describes the action of the continuous translation group $\calT$ on vector fields (as already discussed in the previous section). Hence plane waves are {\it simultaneous eigenfunctions of the continuous translation group $\calT$}. 
\\[2mm]
We now come to the interaction between wave and structure.
The interaction occurs via the generalized structure factor \eqref{genstr} in the diffracted radiation field. Assume for simplicity that the
density is a sum of delta functions at the atom positions, $\rho(y)=\sum_{\bfa\in\calL} \delta_{\bfx_0+\bfa}(y)$. Then this factor is
\beq \label{genstr0}
   \bff_{\bfE_0}(\bfk') = \int_{\R^3} \bfE_0(\bfy)\rho(\bfy)e^{-i\bfk'\cdot\bfy} d\bfy
                        = \sum_{\bfa\in\calL} \bfE_0(\bfx_0+\bfa) e^{-i\bfk'\cdot\bfa}.
\eeq
The eigenfunction property \eqref{DE0} of the waves holds in particular for the
crystalline translations $\bfa\in\calL$, and so it follows that
\beq \label{genstr1}
   \bff_{\bfE_0}(\bfk') = \left(\sum_{\bfa\in\calL} \mbox{(phase factor depending on $\bfa$)} \, e^{-i\bfk'\cdot \bfa}\right) \,  
   \bfE_0(\bfx_0). 
\eeq
Thus the diffraction behaviour is reduced to a {\it phase factor sum}.
The phase factors come from the symmetry of the wave. The points where they are evaluated come from the symmetry of the structure. And the sum ``behaves nicely'':
it interferes constructively when $e^{-i(\bfk'-\bfk_0)\cdot \bfa}=1$ for all $\bfa\in\calL$, i.e. when $\bfk'-\bfk_0$ belongs to the reciprocal lattice $\calL'$, and destructively otherwise.

To get constructive interference, it would be enough if the wave had just {\it the same} symmetry as the crystal, i.e. if \eqref{DE0}--\eqref{phase} was only true for $\bfa$'s in the discrete translation group $\calL$.\footnote{This condition has infinite-dimensionally many solutions: it just means that the electric field multiplied by a complex exponential, $\bfE(\bfx)e^{-i(\bfk'-\bfk_0)\cdot \bfx}$, shares the periodicity of the crystal. In the context of electron wavefunctions instead of electric fields, such waves are known as Bloch waves.} But to get destructive interference when the
incoming radiation parameter $\bfk_0$ or the lattice parameters $A$ are tuned off resonance, one needs \eqref{DE0}--\eqref{phase} for all $\bfa$. 

%In fact, the condition \eqref{DE0}--\eqref{phase} for $\bfa\in\calL$  A simple nearly-periodic field would be a beam with wavelength much larger than the sample diameter, $|\bfk_0|^{-1}=|\bfk'|^{-1}>>1$. Constructive interference would hold regardless of the direction of the incoming beam and the lattice parameters, and hence would not reveal any structural information. 

In summary, the discrete diffraction patterns of classical X-ray crystallography can be traced to the fact that crystals and plane waves have {\it matching symmetries}. 
Constructive interference comes from the fact that plane waves share the symmetry of crystals. Destructive interference comes from the fact that plane waves have a larger, continuous symmetry group.  
\\[2mm]
Everything so far is just an abstract rationalization of a very well known phenomenon. But can it be generalized? 
\\[2mm]
The key is to realize that the foundation on which X-ray crystallography is built, the complex exponential form \eqref{PW0}, can actually be {\it derived} from the innocent looking eigenvalue equation \eqref{DE0}. 
%
%We begin with the structure. The crystal $\calS$ is obtained by translating a single point, 
%$\bfx_0$, by all elements of the lattice $\calL$. 
%Thus $\calS$ is an orbit of a single point, $\bfx_0$, under a discrete group $\calL$ of %translations. The group $\calL$ may be called the ''generating symmetry group'' of the %crystal, and is a subgroup of the full translation group $\calT$.
%
%Now let us look at the wave. Eq. \eqref{DE0} means that $\bfE_0$ is a simultaneous %eigenfunction of the action of all the translations $\bfx_0\mapsto\bfx_0+\bfa$ on 
%vector fields. The special plane-wave form of the eigenvalues \eqref{PHO} can in fact be %{\it derived} from this abstract property. 
If we combine two translations, $\bfx\mapsto(\bfx+\bfa)\mapsto(\bfx+\bfa)+\bfb$, 
we can either apply \eqref{eigen} to the whole translation, or separately to the two
translations by $\bfa$ and $\bfb$, and so 
\beq\label{homo}
   (\mbox{phase factor at }\bfa+\bfb) = (\mbox{phase factor at }\bfa) \cdot (\mbox{phase factor at }\bfb).
\eeq
Mathematically, this means that the phase factor is a {\it group homomorphism} from the translation group to the multiplicative group $\C\backslash\{0\}$. And the only such group homomorphisms are the scalar complex exponentials \eqref{phase}! Eq. \eqref{phase} together with \eqref{DE0} implies \eqref{PW} (with polarization vector $\bfn=\bfE_0(\bfx_0)$), up to the orthogonality constraint on $\bfn$ and $\bfk_0$. The latter follows from 
Maxwell's equations, \eqref{TH}. So plane waves are those solutions to \eqref{DE0} which satisfy Maxwell. (See Theorem \ref{PWthm} for a precise statement.)

Thus we have a path from crystals to plane waves,
\begin{eqnarray}
    \mbox{crystal} & \longrightarrow & \mbox{generating symmetry group} \label{path} \\
                   & \longrightarrow & \mbox{continuous extension of symmetry group}\nonumber \\
                   & \longrightarrow & \mbox{eigenfunctions of continuous symmetry group 
which solve Maxwell}. \nonumber
\end{eqnarray}

This path can be generalized. We can start from any structure which is generated by a symmetry group, but the group does not need to consist of translations. Any discrete subgroup of the euclidean group $E(3)$ is fine. Many interesting
structures in biology and nanotechnology have this form, with the generating subgroup often being a discrete {\it helical} group. The continuous extension of the group is then the helical group $\calH_{\bfe}$ described in eq.~\eqref{hel}. Solving the resulting combined eigenfunction/Maxwell problem will give a different family of incoming waves. 

The non-crystalline but highly symmetric structures which are generated by {\it some} discrete subgroup of $E(3)$ form an interesting class. This class was recently introduced and studied by one of us \cite{James}, and has been named {\it objective structures}. Like crystals, objective structures can be completely classified \cite{JamesDayal}. 

The remarkable constructive/destructive interference properties of the structure factor \eqref{genstr0} survive when structure and radiation are governed by non-translational 
symmetry groups. Here we can rely on the fact that mathematicians such as A. Weil \cite{Weil} realized a long time ago -- albeit with a completely different motivation, number theory rather than molecular biology -- that the Poisson summation formula \eqref{Poisson} has far-reaching generalizations to sums of delta functions over discrete subgroups of continuous groups. For details we refer to Section \ref{sec:TVL}.  
\\[2mm]
To conclude this section, we formulate the design problem suggested by the path \eqref{path} precisely, as a system of equations. As just discussed, crystals can be replaced
by structures generated by a symmetry group which may contain rotations.
\begin{definition} \label{DE} {\bf (Design equations)}. Let $G$ be a closed subgroup of 
the euclidean group $E(3)$ -- the ``desired symmetry of the radiation''.
(Typically, $G$ is obtained as a continuous extension $G\supset G_0$ of a discrete subgroup $G_0$ of $E(3)$ -- the ``generating symmetry of a structure''.) 
A vector field $\bfE_0 \, : \, \R^3\to\C^3$ solves the design equations for $G$ if:
\\[1mm]
(i) $\bfE_0$ is a joint eigenfunction of all group elements $g\in G$; that is to say,
\beq \label{eigen}
           g\,\bfE_0 = \chi(g)\,\bfE_0 \mbox{ for some }\chi(g)\in\C \mbox{ and all }g\in G.
\eeq
Here the left hand side is given by eq. \eqref{action} which describes the action of the euclidean group $E(3)$ on vector fields. 
\\[1mm]
(ii) $\bfE_0$ is a bounded solution to the time-harmonic Maxwell equations
\begin{align} \label{helm}
      & \Delta \bfE_0 = - \frac{\omega^2}{c^2} \bfE_0 \mbox{ for some }\omega>0, \\
      & {\rm div}\, \bfE_0 = 0. \label{div}
\end{align}
\end{definition}
Note that the transformed field $g\bfE_0$ (left hand side in (i)) reduces to the left hand side of \eqref{DE0} when the group element $g$ is given by the translation $(\bfI|-\bfa)$. 

In the next two sections, we will solve the design equations for some interesting examples.
We will make use of the fact that as for \eqref{DE0}, the eigenvalue $\chi(g)$ as a function of $g$ must be a group homomorphism:
\begin{lemma} {\bf (Character lemma).} If $\bfE_0$ is any solution to the design equations
for the group $G$ which is not identically zero, then the function $\chi \, : \, G\to\C$ in Def. \ref{DE} is a bounded continuous group homomorphism from $G$ to 
the multiplicative group $\C\backslash\{0\}$.  
\end{lemma}
{\bf Proof} The argument is the same as that leading to eq. \eqref{homo}. Eq. \eqref{eigen} shows that for $g_1,\,g_2\in G$, 
\beq\label{eigen2}
   \chi(g_1g_2)\bfE_0 = (g_1g_2)\bfE_0 = g_1(g_2\bfE_0) = \chi(g_1) g_2\bfE_0 = 
   \chi(g_1)\chi(g_2)\bfE_0.
\eeq
Evaluation at a point $\bfx$ where $\bfE_0$ is not zero shows that $\chi$ is a group
homomorphism. Boundedness and continuity of $\chi$ are a direct consequence of 
the same properties for $\bfE_0$. 
\\[2mm]
In the case when $G$ is abelian, the bounded continuous group homomorphisms to $\C\backslash\{0\}$ are called the {\it characters} of $G$. The group of characters 
\beq\label{dual}
    G':=\{\chi \, : \, G \to \C\backslash\{0\}\, : \, \chi \mbox{ is a character of }G\}
\eeq
is called the {\it dual group} of $G$, because of the duality relation $(G')'\stackrel{\sim}{=} G$. 

As will become clear after having discussed some examples, the dual group $G'$ can be interpreted physically as a ``wavevector space'' which parametrizes the radiation that solves the design equations, just as the wavevectors $\bfk_0$ in \eqref{PW0} parameterize plane waves.
\\[2mm]
Finally we remark that the design equations as stated in Definition \ref{DE} are intrinsically abelian, and should really only be used for abelian $G$. Namely, \eqref{dual} shows that the action of $G$ on simultaneous eigenfunctions must be abelian, $(g_1g_2)\bfE_0=(g_2g_1)\bfE_0$, since the right hand side in \eqref{eigen2} is independent of the order of the $g_i$. 

For non-abelian but compact subgroups $G$ of $E(3)$,
a generalization of the design equations which can yield radiation families on which $G$ acts in a non-abelian way has been worked out by one of us, and will be presented elsewhere \cite{Juestel}. This may be of interest to analyze structures such as buckyballs and icosahedral viruses \cite{Caspar}, which are generated by non-abelian discrete symmetries. In this case the right notion of characters of $G$ is not given by \eqref{dual}, and has the structure of a hypergroup \cite{Lasser} instead of a group. 
\section{Plane waves as solution to the design equations} \label{sec:PW}
After having formalized our design criterion for structure-adapted radiation into a set of equations (Definition \ref{DE}), we can state our insight from Section \ref{sec:design} that plane waves are right for crystals as a mathematical theorem.
\begin{theorem} \label{PWthm} {\bf (Plane waves are right for crystals).} Let $G$ be the translation group $\calT$ (see \eqref{trans}). Then the solutions to the design equations (Def. \ref{DE}) are precisely the plane waves
\beq \label{PW'}
   \bfE_0(\bfx) = \bfn \, e^{i\bfk_0\cdot\bfx}, \;\; 
   \bfk_0\in\R^3, \;\;\bfn\in\C^3,\;\;\bfk_0\cdot\bfn = 0.
\eeq
\end{theorem}
{\bf Proof} By the character lemma, the function $\chi\, : \, G\to\C$ in \eqref{eigen} must be a character of $\calT$. The characters are well known to be given by 
$$
  \chi_\bfk(\bfa) = e^{i\bfk\cdot\bfa}, \;\;\bfk\in\R^3.
$$
Fix $\bfk_0\in\R^3$ and consider the character $\chi_{-\bfk_0}$. The first eq. ((i) in Def. \ref{DE}) says that 
$$
   \bfE_0(\cdot - \bfa) = e^{i(-\bfk_0)\cdot\bfa}\bfE_0\;\;\mbox{ for all }\bfa\in\R^3.
$$
Evaluation at $\bfx=\bfa$ gives $\bfE_0(0)=e^{-i\bfk_0\cdot\bfx}\bfE_0(\bfx)$, that is to say $\bfE_0$ is of form $\bfn e^{i\bfk_0\cdot\bfx}$ with $\bfn=\bfE_0(0)$. The first of the Maxwell equations in Def. \ref{DE} holds automatically. The second one holds if and only if $\bfk_0\cdot\bfE_0(0)=0$. This completes the proof.
\\[2mm]
The proof says that fixing a wavevector $\bfk_0$ corresponds precisely to fixing a character $\chi\, : \, G\to\C$ in the symmetry condition \eqref{eigen}. For each fixed character, the solutions of the design equations form a complex vector space parametrized by $\{\bfn\in\C^3 \, : \, \bfk_0\cdot\bfn=0\}$. This vector space has dimension 2, except in the special case $\chi=1$, where the dimension is $3$.
\\[2mm]
Let us also look at what happens in the closely related case when $G$ is a {\it two-dimensional} translation group, 
\beq \label{trans2}
   \calT_{\bfe^\perp} = \{ (\bfI|\bfa)\, : \, \bfa \in\R^3, \; \bfa\cdot\bfe = 0\}.
\eeq
Here $\bfe$ is a given unit vector in $\R^3$. This group naturally arises as the continuous extension of the discrete generating symmetry group $G_0$ of a 2D crystalline sheet, an important example being graphene.
For the group \eqref{trans}, we claim that the solutions to the design equations (Def. \ref{DE}) are the plane-wave pairs
\beq \label{sol2D}
   \bfE_0(\bfx) = \bfn_+ e^{i\bfk_0\cdot\bfx} + \bfn_- e^{i(\bfI-2\bfe\otimes\bfe)\bfk_0 
                 \cdot x}, \;\; \bfn_{\pm}\in\C^3, \;\; \bfk_0\cdot\bfn_+=0, \;\;
                 (\bfI-2\bfe\otimes\bfe)\bfk_0\cdot\bfn_-=0,
\eeq
with the corresponding frequency in the design equations given by $\omega=c|\bfk_0|$. 
Note that the two wavevectors $\bfk_0$ and  $(\bfI-2\bfe\otimes\bfe)\bfk_0$ are the mirror images of each other with respect to the symmetry plane $\bfk\cdot\bfe=0$.

We sketch the argument leading to \eqref{sol2D}. We may assume $\bfe=(0,0,1)$. The eigenfunction property under the symmetry gives  $\bfE_0(x_1,x_2,x_3)=e^{i(k_1x_1+k_2x_2)}\bfE_0(0,0,x_3)$ for some planar wavevector 
$(k_1,k_2)\in\R^2$. The fact that $\bfE_0$ must be a bounded solution to Maxwell's equations then gives that the dependence on $x_3$ is also of complex exponential form, 
$\bfE_0(0,0,x_3)=\bfn_+ e^{ik_3x_3} + \bfn_- e^{-ik_3x_3}$. The remaining assertions are then straightforward.
\section{Twisted waves} \label{sec:TW}
We now look at the case when $G$ is the helical group $\calH_{\bfe}$, \eqref{hel}. This group naturally arises as the continuous extension of any discrete generating symmetry group of a helix or nanotube structure. 

We parametrize the group by the rotation angle $\theta$ about the helical axis $\bfe$, 
\beq \label{par}
   \calH_{\bfe} = \{(\bfR_\theta|\tau\bfe)\, : \, \theta\in[0,2\pi),\,\tau\in\R\}, \;\;
   \bfR_\theta = \begin{pmatrix} \cos\theta & -\sin\theta & 0 \\
                              \sin\theta &  \cos\theta & 0 \\
                              0          &       0     & 1 \end{pmatrix}.
\eeq
It will be convenient to work in cylindrical coordinates with respect to the helical axis, that is to say 
\beq \label{cyl}
  \bfe = \bfe_3 = \begin{pmatrix} 0 \\ 0 \\ 1 \end{pmatrix}, \;\;
  \bfx = \begin{pmatrix} x_1 \\ x_2 \\ x_3 \end{pmatrix} =
  \begin{pmatrix} r\cos\varphi \\ r\sin\varphi \\ z \end{pmatrix}, \;\; r\in[0,\infty), 
  \, \varphi\in[0,2\pi),\, z\in\R.
\eeq
In cylindrical coordinates, the action \eqref{action} of the helical group on vector fields assumes the following simple form:
\beq\label{action2}
   \Bigl( (\bfR_\theta|\tau\bfe)\bfE \Bigr)(r,\varphi,z) = \bfR_\theta  
   \bfE(r,\varphi-\theta,z-\tau).
\eeq
See Figure 3.
\begin{figure}[http!] 
\begin{center}
\includegraphics[width=0.5\textwidth]{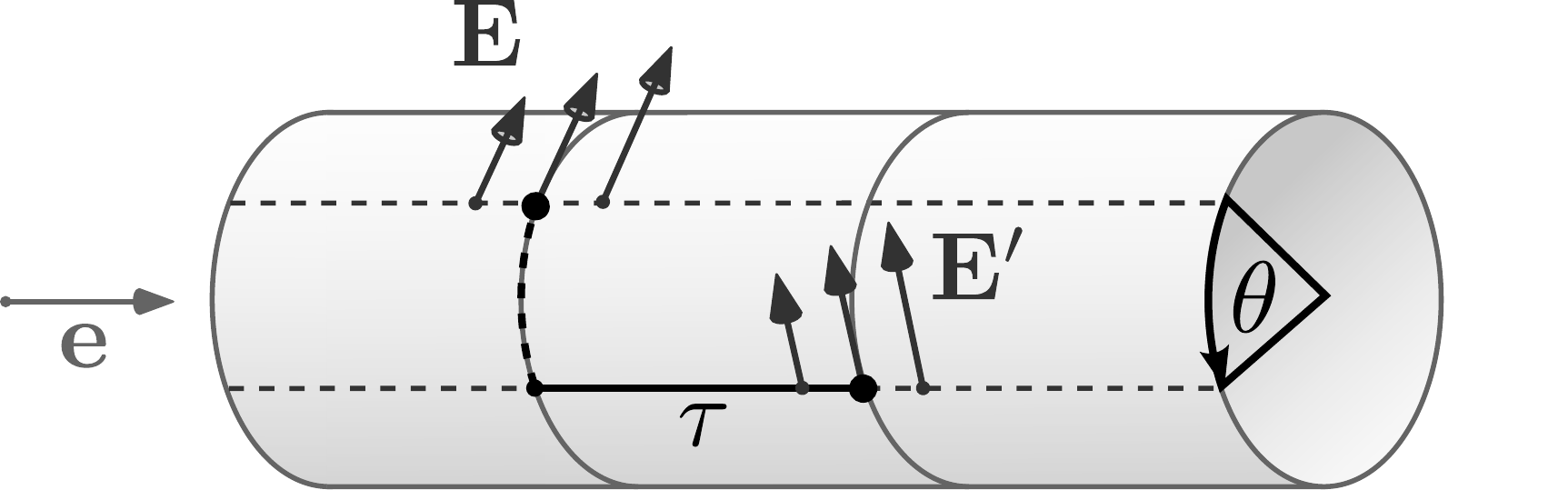}
\caption{{\it Action of the helical group on vector fields and design of twisted waves.} Start from a given vector field $\bfE$. The transformed field $\bfE'$ (right hand side of eq.~\eqref{action2}) under the action of a typical element of the helical group is obtained as follows:
(i) translate the base points along the helical axis $\bfe$ by some amount $\tau$ while leaving the field direction unchanged; (ii) rotate the base points by some angle $\theta$; and (iii) rotate the field direction by the same angle. The design equations require that the transformed field only differs from the original field by a phase factor, regardless of how $\tau$ and $\theta$ are chosen. Time-harmonic electric fields with this property are named {\it twisted waves}, and can be proven to have the mathematical form \eqref{TW} (see Theorem \ref{TWthm}).}
\end{center}
\label{fig:action}
\end{figure}
Here and below, $\bfE$ is a function from polar coordinate space to the cartesian space $\R^3$, that is to say $E_1$, $E_2$, $E_3$ are the cartesian field components of $\bfE$ and the action of the group on the direction of the field vectors is the usual action of the 3$\times$3 matrix $\bfR_\theta$ on vectors. 
\begin{theorem} \label{TWthm} {\bf (Twisted waves).} Let $G$ be the helical group $\calH_{\bfe}$ with axis $\bfe$. Then the solutions to the design equations (Def. \ref{DE}) are precisely 
\beq \label{TW}
  \bfE_0(r,\varphi,z) = e^{i(\alpha\varphi+\beta z)} \, \bfR_\varphi \,  \bfN(\bfn)  
  \begin{pmatrix}  \Jp(\gamma r) \\ \Jm(\gamma r) \\ \J(\gamma r)
  \end{pmatrix},
\eeq
with $(\alpha,\beta,\gamma)\in\Z\times\R\times(0,\infty)$ (``parameter vector''), $\bfn\in\C^3$ (``polarization vector''), and 
\beq \label{divfree}
           (0,\gamma,\beta)\cdot \bfn = 0. 
\eeq
Here $\bfN(\bfn)$ is a certain 3$\times$3 matrix (``polarization tensor'') which depends linearly on the polarization vector $\bfn$, 
\beq \label{N}
     \bfN(\bfn) = \begin{pmatrix}  
 \mbox{$\frac{n_1+in_2}{2}$} & \mbox{$\frac{n_1-in_2}{2}$} & 0 \\
 \mbox{$\frac{n_2-in_1}{2}$} & \mbox{$\frac{n_2+in_1}{2}$} & 0 \\
  0 & 0 & n_3     
     \end{pmatrix},
\eeq 
the $\J$ are Bessel functions, $(r,\varphi,z)$ are cylindrical coordinates \eqref{cyl} with respect to the helical axis, and $\bfE_0$ is the cartesian field vector. The associated frequency in the design equations is given by $\omega=c|(0,\gamma,\beta)|$. 
\end{theorem}
We call the electric fields \eqref{TW}--\eqref{N} {\it twisted waves}. 
Figure \ref{fig:TW} in the Introduction shows the twisted wave with parameter vector $(\alpha,\beta,\gamma)=(5,3,1)$ and polarization vector $\bfn=(1,0,0)$. 

As already remarked in the Introduction, twisted waves consist of four factors: 
a scalar plane wave on the cylinder; a rotation matrix which rotates the field direction along with the base point; a somewhat mysterious polarization tensor; 
and a vector of three Bessel functions of neighbouring order. 

A subtle aspect is that each twisted wave possesses {\it two} analoga of the plane-wave wavevector $\bfk_0$ in \eqref{PW'}: first, the parameter vector $(\alpha,\beta,\gamma)$ which parametrizes the radiation and consists of an angular, an axial and a radial wavenumber; and second, the cartesian vector $(0,\gamma,\beta)$ which determines the allowed plane of polarization vectors as well as the frequency $\omega$ of the wave (or, equivalently, its wavelength, see eq. \eqref{wavelength}). The latter vector has a simple but non-obvious physical meaning which will emerge in Section \ref{sec:Fourier}.    

The parameters $\alpha$ and $\beta$ can be interpreted as eigenvalues of angular momentum respectively momentum. Namely, it is easily checked that the twisted wave \eqref{TW}, \eqref{N} is an exact solution to the eigenvalue equations
\begin{equation} \label{eval}
  J_z \bfE_0 = \alpha\,\bfE_0, \;\; P_z \bfE_0 = \beta\, \bfE_0, \;\; \mbox{ where }
  J_z = \frac{1}{i} \frac{\partial}{\partial\varphi} + 
  \mbox{$\begin{pmatrix} 0 & -i & \\ i & 0 & \\ & & 0 \end{pmatrix}$}, \;\; P_z = \frac{1}{i}
  \frac{\partial}{\partial z}.
\end{equation}
Here $P_z$ is the well known quantum mechanical momentum operator in axial direction, and $J_z$ is the correctly defined angular momentum operator on vector fields with respect to the helical axis (whose cartesian form \eqref{infgen1} can be found in \cite{CT}). These operators arise in our context of classical electrodynamics as the infinitesimal generators of the action \eqref{action} of the rotational and translational subgroup of the helical group \eqref{par} on vector fields: in  cartesian coordinates,  
\begin{eqnarray} 
  & & \frac{d}{d\theta} \bfR_\theta \bfE(\bfR_\theta^{-1}\bfx)\Big|_{\theta=0} = 
      \Bigl( - \bfe\cdot (\bfx\wedge\nabla) + \bfe\wedge\Bigr) \bfE(\bfx) = 
      \frac{1}{i} (J_z\bfE)(\bfx), \label{infgen1} \\
  & & \frac{d}{d\tau} \bfE(\bfx-\tau\bfe)\Big|_{\tau=0} = - (\bfe\cdot\nabla)\bfE(\bfx) = \frac{1}{i}(P_z\bfE)(\bfx).
\end{eqnarray}
We remark that the eigenvalue equations \eqref{eval} are equivalent to the design equation \eqref{eigen} with character $\chi(\theta,\tau)=e^{-i(\alpha\theta + \beta\tau)}$. 
\\[2mm]
{\bf Proof} We split the proof into three parts, dealing in turn with the three conditions in Def. \ref{DE}: the symmetry condition, the Helmholtz equation, and the divergence condition. 
\\[1mm]
{\bf Step 1: Symmetry condition.} By the character lemma, the function $\chi$ in \eqref{eigen} must be a character of the helical group $\calH_{\bfe}$. The parametrization 
\eqref{par} shows that the helical group $\calH_\bfe$ is isomorphic to $S^1\times\R$, where $S^1 \stackrel{\sim}{=} [0,2\pi)$ with the usual addition of angles modulo $2\pi$.
The characters $\chi\, : \, S^1\times\R\to\C\backslash\{0\}$ of $S^1\times\R$ are well known to be 
$$
   \chi_{\alpha,\beta}(\theta,\tau) = e^{i(\alpha\theta + \beta\tau)}, 
   \;\;(\alpha,\beta)\in\Z\times\R.
$$
In particular, the dual group \eqref{dual} is given by 
$$
   (\calH_{\bfe})' \stackrel{\sim}{=} \Z\times\R.
$$
Fix $(\alpha,\beta)\in\Z\times\R$ and consider the character $\chi_{-(\alpha,\beta)}\, : \, S^1\times\R\to\C\backslash\{0\}$. The first design equation ((i) in Def. \ref{DE}) says that 
$$
   \bfR_\theta \bfE_0(r,\varphi-\theta,z-\tau) =
        e^{i(-\alpha\varphi-\beta\tau)}\bfE_0(r,\varphi,z).
$$
Evaluation at $\varphi=\theta$, $z=\tau$ gives 
\beq \label{anstw}
   \bfE_0(r,\varphi,z) = e^{i(\alpha\varphi+\beta z)} \bfR_\varphi \bfE_0(r,0,0).
\eeq
{\bf Step 2: Helmholtz equation.} Substituting the ansatz \eqref{anstw} into the Helmholtz equation (first of the Maxwell equations in Def. \ref{DE}) gives an ordinary differential equation for $\bfE_0(r,0,0)$. But unlike in the case of the 2D translation group $\calT_{\bfe^\perp}$, due to the presence of the rotation matrix $\bfR_\varphi$ this ODE will not decouple into independent ODE's for the components. 

The idea to overcome this difficulty is to simultaneously diagonalize the matrices $\bfR_\varphi$, $\varphi\in[0,2\pi)$, with a unitary transformation. Simultaneous diagonalization is possible because the group $SO(2)$ of these matrices is {\it abelian}. We have
\beq \label{simdiag}
   \bfR_\varphi = U 
   \begin{pmatrix} 
      \mbox{$e^{i\varphi}$} & & \\ & \mbox{$e^{-i\varphi}$} & \\ & & 1  
   \end{pmatrix} 
   U^{-1} \mbox{ for all }\varphi\in[0,2\pi), \mbox{ with } U =
   \begin{pmatrix}  
      \mbox{$\frac{i}{\sqrt{2}}$} & \mbox{$\frac{-i}{\sqrt{2}}$} & 0 \\
      \mbox{$\frac{1}{\sqrt{2}}$} & \mbox{$\frac{1}{\sqrt{2}}$} & 0 \\
       0  &  0  &  1
   \end{pmatrix}.
\eeq    
Let $\tilde{\bfE}(r,\varphi,z):=U^{-1}\bfE_0(r,\varphi,z)$. By \eqref{anstw} and \eqref{simdiag}, we have
\beq \label{Etilde2}
   \tilde{\bfE}(r,\varphi,z) = e^{i(\alpha\varphi + \beta z)} 
   \begin{pmatrix} e^{i\varphi} & & \\ & e^{-i\varphi} & \\ & & 1 \end{pmatrix} 
   \tilde{\bfE}(r,0,0).
\eeq
Since the Helmholtz equation is invariant under the transformation $\bfE_0\mapsto\tilde{\bfE}=U^{-1}\bfE_0$, we can substitute $\tilde{\bfE}$ into this eq. and obtain, using that the Laplacian in cylindrical coordinates is
$$
   \Delta = \frac{\partial^2}{\partial r^2} + \frac{1}{r}\frac{\partial}{\partial r} + 
   \frac{\partial^2}{\partial\varphi^2} + \frac{\partial^2}{\partial z^2},
$$
the following ODE for the components: 
\beq \label{ODEtw}
   \left( \frac{\partial^2}{\partial r^2} + \frac{1}{r}\frac{\partial}{\partial r} - 
   \frac{(\alpha+\sigma_j)^2}{r^2} + \left( (\mbox{$\frac{\omega}{c}$})^2 - \beta^2\right) \right) \tilde{E}_j(r,0,0) = 0, \;\; j=1,2,3,
\eeq
where $\sigma_1=1, \, \sigma_2=-1, \, \sigma_3=0$. That is to say, the radial functions $\tilde{E}_j(r,0,0)$ are solutions to Bessel's equation. Boundedness of $\tilde{E}_j$ implies that we must have $(\omega/c)^2-\beta^2\ge 0$. Moreover, for integer values of $\alpha$ there is only a one-dimensional space of bounded solutions, given by
\beq\label{Bessel}
  \tilde{E}_j(r,0,0) = c_j J_{\alpha+\sigma_j}(\gamma r), \;\; \gamma = \sqrt{(\mbox{$\frac{\omega}{c}$})^2-\beta^2}, \;\; c_j\in\C.
\eeq
By applying $U$ we obtain the original field $\bfE_0(r,0,0)$, 
$$
   \bfE_0(r,0,0) = U\, {\rm diag}(\bfc) 
   \begin{pmatrix}  \Jp(\gamma r) \\ \Jm(\gamma r) \\ \J(\gamma r)\end{pmatrix}, \;\; \bfc\in\C^3,
$$
with $\gamma$ as in \eqref{Bessel}. For reasons that are not apparent at this point but will emerge later, it is fruitful to parametrize the solution space not by $\bfc\in\C^3$ but by
\beq \label{n}
   \bfn := U\bfc  =  \begin{pmatrix} i\frac{c_1-c_2}{\sqrt{2}} \\ 
                                      \frac{c_1+c_2}{\sqrt{2}} \\
                                             c_3 \end{pmatrix} \in \C^3.
\eeq
It follows that 
\beq \label{startvector}
   \bfE_0(r,0,0) = \bfN(\bfn) 
   \begin{pmatrix}  \Jp(\gamma r) \\ \Jm(\gamma r) \\ \J(\gamma r)\end{pmatrix}, \;\;
   \bfn\in\C^3, \;\; \bfN(\bfn):=U \, {\rm diag}(U^{-1}\bfn).
\eeq
Here and below, ${\rm diag}(\bfc)$ denotes the diagonal matrix whose diagonal entries are given by the components of the vector $\bfc$. Using the explicit form of $U$ (see eq. \eqref{simdiag}) it is easy to check that the
$3\times 3$ matrix $\bfN(\bfn)$ introduced above is given by the expression in the theorem. Substitution into \eqref{anstw} shows that $\bfE_0$ has the form \eqref{TW}, except that $\bfn\in\C^3$ is still arbitrary.
\\[2mm]
{\bf Step 3: Divergence condition.} It remains to analyze the second Maxwell equation, ${\rm div}\, \bfE_0=0$. This is not straightforward, due to the fact that the field components are cartesian but the coordinates are cylindrical. We begin by expressing the field components in cylindrical coordinates also. 

We denote the unit vectors in the direction of $r$, $\varphi$ and $z$ by $\bfe_r$, $\bfe_\varphi$, $\bfe_z$, that is to say
\begin{equation} \label{erephiez}
  \bfe_r = \begin{pmatrix}\cos\varphi \\ \sin\varphi \\ 0 \end{pmatrix}, \;\;    
  \bfe_\varphi = \begin{pmatrix}-\sin\varphi \\ \cos\varphi \\ 0 \end{pmatrix}, \;\;
  \bfe_z = \begin{pmatrix} 0 \\ 0 \\ 1 \end{pmatrix}.
\end{equation}
The cylindrical field components are $E_r=\bfe_r\cdot\bfE_0$, $E_\varphi=\bfe_\varphi\cdot\bfE_0$, $E_z=\bfe_z\cdot \bfE_0$. Since the cylindrical unit vectors are the image of the cartesian unit vectors under the rotation $\bfR_\varphi$, 
i.e. $\bfe_r=\bfR_\varphi\bfe_1$, $\bfe_\varphi=\bfR_\varphi\bfe_2£$, $\bfe_z=\bfR_\varphi\bfe_3$, the rotation matrix cancels from the cylindrical field components:
$$
 E_r = e^{i(\alpha\varphi+\beta z)} \bfE_0(r,0,0)\cdot \bfe_1, \;\;\,
 E_\varphi = e^{i(\alpha\varphi+\beta z)} \bfE_0(r,0,0)\cdot \bfe_2, \;\;\,
 E_z = e^{i(\alpha\varphi+\beta z)} \bfE_0(r,0,0)\cdot \bfe_3.
$$
Moreover, in terms of $\bfc=U^{-1}\bfn$ the polarization tensor is
\beq \label{polc}
   \bfN(\bfn) = \begin{pmatrix}
   \mbox{$\frac{ic_1}{\sqrt{2}}$} & \mbox{$\frac{-ic_2}{\sqrt{2}}$} & 0 \\
      \mbox{$\frac{c_1}{\sqrt{2}}$} & \mbox{$\frac{c_2}{\sqrt{2}}$} & 0 \\
       0  &  0  &  c_3
   \end{pmatrix}.
\eeq
It follows that 
\beq \label{TWcyl}
    \begin{pmatrix} E_r \\ E_\varphi \\ E_z \end{pmatrix}(r,\varphi,z) = 
    e^{i(\alpha\varphi+\beta z)} \bfN(\bfn) 
    \begin{pmatrix} \Jp(\gamma r) \\ \Jm(\gamma r) \\ \J(\gamma r) \end{pmatrix} =
    e^{i(\alpha\varphi+\beta z)}
    \begin{pmatrix} \mbox{$\frac{i}{\sqrt{2}}$} (c_1\Jp(\gamma r) - c_2\Jm(\gamma r)) \\
    \mbox{$\frac{1}{\sqrt{2}}$} (c_1\Jp(\gamma r) + c_2\Jm(\gamma r)) \\
    c_3\J(\gamma r) \end{pmatrix}.
\eeq
Applying the formula for the divergence of a vector field $\bfv$ in cylindrical coordinates,
\beq \label{divcyl}
  {\rm div}\, (v_r\bfe_r + v_\varphi \bfe_\varphi + v_z\bfe_z) = 
  \frac{\partial v_r}{\partial r} + \frac{1}{r} v_r  + \frac{1}{r} \frac{\partial v_\varphi}{\partial\varphi} + \frac{\partial v_z}{\partial z},
\eeq
gives 
$$
  {\rm div}\, \bfE_0 \! = \!  e^{i(\alpha\varphi+\beta z)}
  \left\{\! 
  \tfrac{i}{\sqrt{2}}\left( c_1\gamma \Jp' \! - c_2 \gamma \Jm' \! + \frac{c_1}{r}\Jp 
  \! - \frac{c_2}{r}\Jm\right) + \tfrac{1}{\sqrt{2}}\left( \frac{i\alpha c_1}{r}\Jp \! + 
  \frac{i\alpha c_2}{r}\Jm\right) + i\beta c_3\J
  \! \right\}.
$$
Here we have dropped the argument $\gamma r$ of the Bessel functions.
By Bessel's identities 
\beq \label{Besselid}
  \Jp'(z) = \J(z) - \frac{\alpha\!+\! 1}{z}\Jp(z), \;\;\;\;
  \Jm'(z) = -\J(z) + \frac{\alpha\!-\! 1}{z}\Jm(z),  
\eeq
the terms involving $\Jp$ and $\Jm$ cancel and we obtain the final result
\beq\label{divE0}
  {\rm div}\, \bfE_0 =  i \left(\gamma\frac{c_1+c_2}{\sqrt{2}} + \beta c_3\right)
  \, e^{i(\alpha\varphi+\beta z)}  \J(\gamma r) = i \left((0,\gamma,\beta)\cdot \bfn\right) \, e^{i(\alpha\varphi+\beta z)}
  \J(\gamma r).
\eeq
In the last equation we have re-expressed the vector $\bfc$ in terms of $\bfn$. In particular, we see that the field is divergence-free if and only if eq. \eqref{divfree} holds. This completes the proof of Theorem \ref{TWthm}.
\\[2mm]
Next we present an interesting algebraic property of the somewhat mysterious polarization tensor which emerged from the above proof.
\begin{lemma} {\bf (Intertwining lemma).} The polarization tensor $\bfN(\bfn)$ introduced in \eqref{startvector} satisfies 
\beq \label{twine}
   \bfR_\varphi \bfN(\bfn) = \bfN(\bfn) 
   \begin{pmatrix} e^{i\varphi} & & \\ & e^{-i\varphi} & \\ & & 1 \end{pmatrix}, 
   \;\mbox{ for all }\bfn\in\C^3.
\eeq
\end{lemma}
Eq. \eqref{twine} means that $\bfN(\bfn)$ ``intertwines'' the standard representation and the diagonal representation of the rotational subgroup $SO(2)$ of the helical group $\calH_{\bfe}$ on $\C^3$.
\\[2mm] 
{\bf Proof} By the diagonal representation \eqref{simdiag} of $\bfR_\varphi$ and the fact that diagonal matrices commute, we have, abbreviating the diagonal matrix in the lemma by $D_\varphi$, 
$$
   \bfR_\varphi\bfN(\bfn) = U\, D_\varphi U^{-1} \bfN(\bfn) = U\, D_\varphi {\rm diag}(U^{-1}\bfn) = U \, {\rm diag}(U^{-1}\bfn) D_\varphi = \bfN(\bfn)\, D_\varphi.
$$ 
The lemma immediately implies an equivalent and sometimes useful representation of the twisted wave \eqref{TW} which we note for future reference:
\beq \label{TW'}
  \bfE_0(r,\varphi,z) = e^{i(\alpha\varphi+\beta z)}\bfN(\bfn) 
  \begin{pmatrix} e^{i\varphi} & & \\ & e^{-i\varphi} & \\ & & 1 \end{pmatrix} 
  \begin{pmatrix} \Jp(\gamma r) \\ \Jm(\gamma r) \\ \J(\gamma r) \end{pmatrix}.
\eeq
Finally we compute the magnetic field associated to a twisted wave. We use the  cylindrical components \eqref{TWcyl} of the wave and apply the formula for the curl of a vector field $\bfv$ in cylindrical coordinates:
$$
   {\rm curl} \, (v_r\bfe_r + v_\varphi \bfe_\varphi + v_z \bfe_z) = 
   \begin{vmatrix} \frac{1}{r}\bfe_r & \bfe_\varphi & \frac{1}{r}\bfe_z \\
   \frac{\partial}{\partial r} & \frac{\partial}{\partial \varphi} &
              \frac{\partial}{\partial z} \\
   v_r & r\, v_\varphi & v_z  \end{vmatrix}.
$$ 
After some calculation, we find using \eqref{Besselid} that 
\beq \label{TWcurl}
  {\rm curl} \, \bfE_0 = e^{i(\alpha\varphi+\beta z)} 
  \bfR_\varphi \bfN(i\bfk_0\times \bfn) 
  \begin{pmatrix} \Jp(\gamma r) \\ \Jm(\gamma r) \\ \J(\gamma r) \end{pmatrix}, \;\;
  \bfk_0 := \begin{pmatrix} 0 \\ \gamma \\ \beta \end{pmatrix}. 
\eeq
That is to say, to obtain the curl one just has to replace the vector $\bfn$ inside the polarization tensor by $i\bfk_0\times\bfn$. 

The magnetic field associated with the twisted wave \eqref{TW} can be immediately read off from \eqref{B0} and \eqref{TWcurl}:
\beq \label{TWmagn}
   \bfB_0(r,\varphi,z) = e^{i(\alpha\varphi+\beta z)} \, \bfR_\varphi \,  \bfN(\mbox{$\frac{1}{\omega}$} \bfk_0\times \bfn)  
  \begin{pmatrix}  \Jp(\gamma r) \\ \Jm(\gamma r) \\ \J(\gamma r)
  \end{pmatrix}, \;\;\; \mbox{ with }\bfk_0\mbox{ as in \eqref{TWcurl}}.
\eeq
Hence for twisted waves, with the ``right'' definition of wavevector the map from $\bfE_0$ to $\bfB_0$ is precisely the same map on the polarization vector $\bfn$ as for plane waves \eqref{PW}. A deeper understanding of this fact will be achieved in Section \ref{sec:Fourier}.
\section{The reciprocal lattice of a helical structure} \label{sec:reciprocal}
Crystals are orbits of a finite set of atoms under a discrete group of translations (a lattice). Helical structures are orbits of a finite set of atoms under a discrete subgroup of the helical group. For the latter structures, in the context of X-ray fiber diffraction a notion of ``reciprocal helical lattice'' has been introduced by Klug, Crick, and Wyckham \cite{KlugEtAl}, by periodically extending the helical subgroup to a Bravais lattice in $\R^2$ and applying the concept of reciprocal Bravais lattice. 
We show here that this notion of reciprocal helical lattice has an {\it intrinsic group-theoretic meaning} which parallels, rather than needs to rely on, that of the reciprocal lattice in the crystal case.
This group-theoretic meaning will be very helpful in understanding and interpreting the diffraction patterns of helical structures subjected to {\it twisted waves}. 

Recall from \eqref{pos1} that the atomic positions in a crystal structure, 
$$
   {\cal S} = \{\bfx_0^{(\nu)} + a \, : \, a\in\calL, \, \nu=1,..,M\},
$$
are the orbit of a finite set of positions $\bfx_0^{(1)},..,\bfx_0^{(M)}$ under a Bravais lattice $\calL$ (i.e., a discrete subgroup of the translation group of form $\calL=A\,\Z^3$ for some invertible 3$\times$3 matrix $A$). Analogously, the atomic positions in a helical structure,
\begin{equation} \label{He}
   {\cal S} = \{g \, \bfx_0^{(\nu)}  \, : \, g\in H_0, \, \nu=1,..,M\},
\end{equation}
are the orbit of a finit set of points $\bfx_0^{(1)},..,\bfx_0^{(M)}$ under a {\it discrete helical group}, by which we mean a discrete subgroup of the helical group $\calH_{\bfe}$ with axis $\bfe$ (see \eqref{hel}) of form
\begin{equation}\label{H0}
  H_0 = \{ g_0^i h_0^j \, : \, i\in\Z, \, j=1,..,n\}, \; 
  h_0 = (\bfR_{\theta_0}|\tau_0\bfe), \; g_0 = (\bfR_{2\pi/n}|0), \; \theta_0\in[0,2\pi), \, 
  \tau_0\in\R, \, \tau_0\neq 0, \, n\in\N. 
\end{equation}
In case $n=1$, $g_0$ is the identity and $\calS$ is a helix. In case $n>1$, $\calS$ is the union of $n$ helices, basic examples being ``zigzag'' or ``armchair'' carbon nanotubes. The parameters $\tau_0$ and $\theta_0$ of the generating screw displacement $h_0$ encode the pitch $p$ and the number $q$ of subunits per turn of the helices. The pitch is defined as the axial displacement for one full rotation, 
\beq \label{pitch}
    p = \tau_0 \, \frac{2\pi}{\theta_0},
\eeq
and the number of subunits (i.e. rotated and translated copies of the set
$\{\bfx_0^{(1)},..,\bfx_0^{(M)}\}$) per turn is
\beq \label{NperTurn}
    q = \frac{2\pi}{\theta_0}.
\eeq
In physical and biological examples, the latter number is typically a rational number but not an integer.

\begin{figure}[http!]  
\begin{center}
\includegraphics[width=0.42\textwidth]{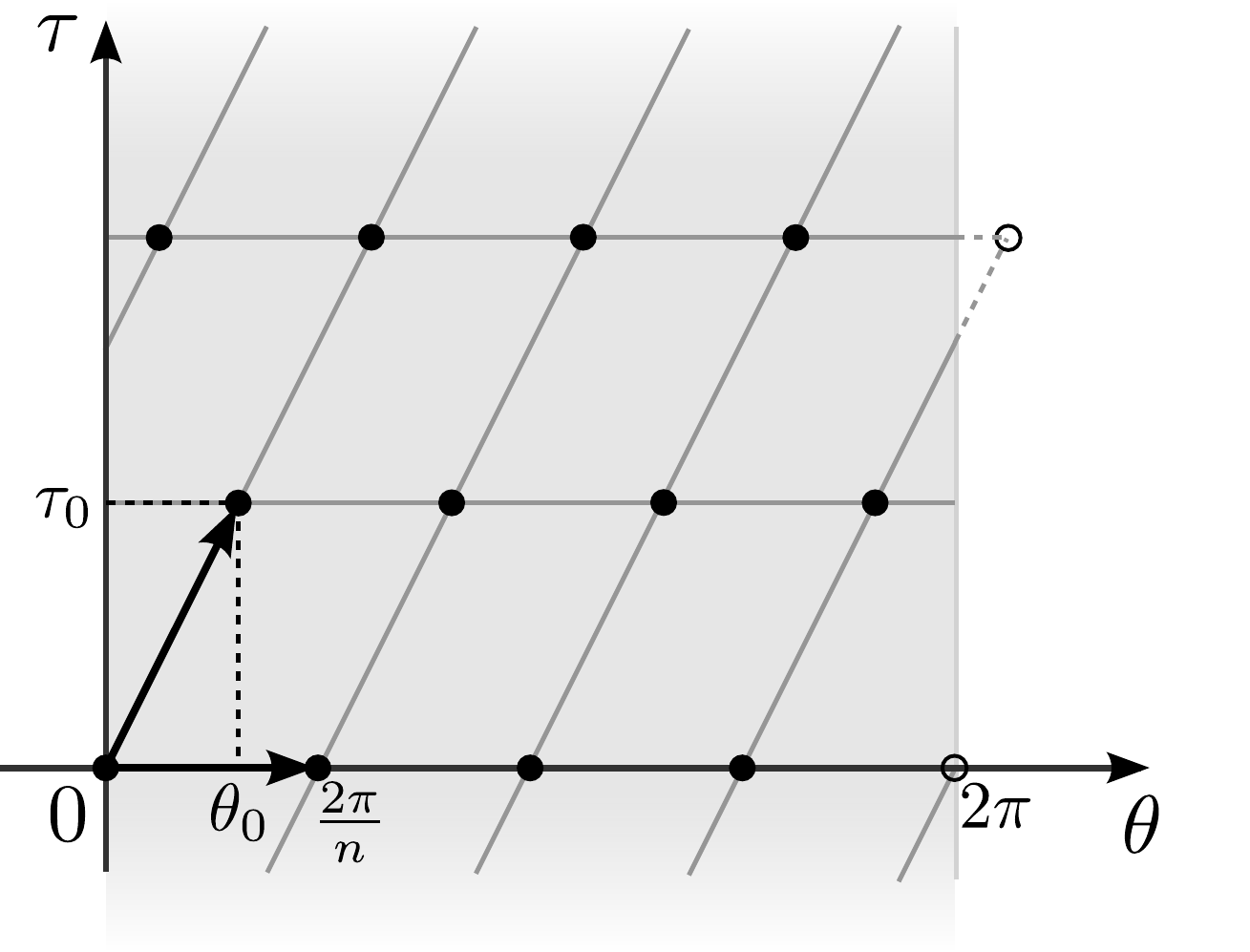} \hspace*{2mm}
\includegraphics[width=0.54\textwidth]{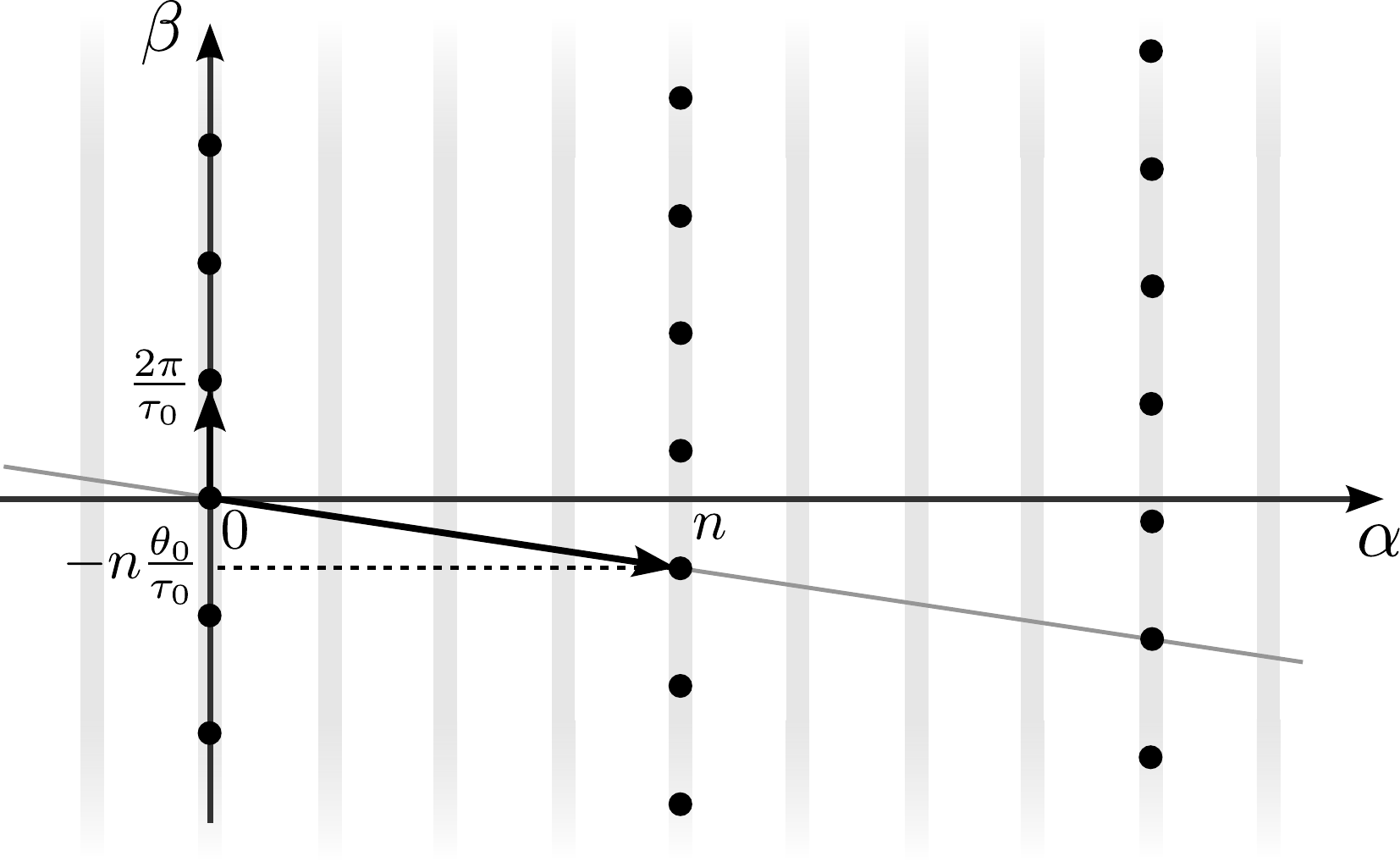}
\end{center}
\caption{A discrete helical group (left, eq. \eqref{H02}) and the associated reciprocal helical lattice (right, eqs.~\eqref{H0'}, \eqref{H0'expl}). Here $\theta$ is the rotation angle about the helical axis and $\tau$ the displacement along the axis, and the reciprocal parameters $\alpha$ and $\beta$ are an angular respectively axial wavenumber. The slope of the reciprocal basis vector pointing to the right is the inverse pitch multiplied by $2\pi$, by eq.~\eqref{pitch}. The discrete helical group is a subset of the full helical group ${\calH_\bfe}\tilde{=}S^1\!\times\R$ (eq.~\eqref{par}, shaded vertical strip), whereas the reciprocal helical lattice is a subset of the dual group $\Z\times\R$ (eq.~\eqref{H'}, shaded vertical lines).}
\label{fig:H0'}
\end{figure}

Before giving explicit examples, we bring formula \eqref{H0} into a form more similar to a Bravais lattice, by writing the group elements in the form $(\bfR_\theta|\tau\bfe)$. We have
$$
   g_0^jh_0^j = (\bfR_\theta|\tau\bfe) \mbox{ with }\theta=\frac{2\pi}{n}i + \theta_0 j \,\mbox{mod}\, 2\pi, \, \tau=\tau_0 j.
$$
(Here $\tilde{\theta}\,\mbox{mod}\, 2\pi$ denotes the unique angle in $[0,2\pi)$ which differs from $\tilde{\theta}$ by an integer multiple of $2\pi$.) The above equations for $\theta$ and $\tau$ can be written more compactly as
\begin{equation} \label{H0el}
  \begin{pmatrix} \theta \\ \tau \end{pmatrix} = 
  i \begin{pmatrix} \frac{2\pi}{n} \\ 0 \end{pmatrix} +
  j \begin{pmatrix} \theta_0 \\ \tau_0 \end{pmatrix} \,\mbox{mod}\,
  \begin{pmatrix} 2\pi \\ 0 \end{pmatrix} =
  \begin{pmatrix} \frac{2\pi}{n} & \theta_0 \\
                  0 & \tau_0 \end{pmatrix} 
   \begin{pmatrix} i \\ j \end{pmatrix}
   \,\mbox{mod}\,
  \begin{pmatrix} 2\pi \\ 0 \end{pmatrix}.
\end{equation}
Denoting this parameter set by $\calH_0$, i.e.
\beq \label{calH0}
    \calH_0 = A\,\Z_n\times\Z \,\mbox{mod}\, \begin{pmatrix} 2\pi \\ 0 \end{pmatrix}
    \; \mbox{ with } A = \begin{pmatrix} \frac{2\pi}{n} & \theta_0 \\
                  0 & \tau_0 \end{pmatrix},
\eeq
it follows that 
\begin{equation} \label{H02}
   H_0  = \{(\bfR_\theta|\tau\bfe) \, : \, \begin{pmatrix} \theta \\ \tau \end{pmatrix}
   \in \calH_0\}, \;  
   \theta_0\in[0,2\pi), \, \tau_0\in\R, \, \tau_0\neq 0, \, n\in\N.
\end{equation}
(Here, $v\,\mbox{mod}(2\pi,0)$ denotes the unique vector in $[0,2\pi)\times\R$ which differs from $v$ by an integer multiple of $(2\pi,0)$.) See Figure \ref{fig:H0'}. 
\\[2mm]
{\bf Example 1: Carbon nanotubes.} All nanotubes are of form \eqref{He}, \eqref{H02}. To give a specific example, single-walled (6,5) Carbon nanotubes with axis $\bfe=(0,0,1)$ correspond to 
\begin{eqnarray*}
   M &=& 2 \; \mbox{(number of atoms per unit cell)} \\
   n &=& 1 \; \mbox{(single helix)} \\
   \theta_0 &=& \frac{149}{182} \, 2\pi \\
   \tau_0 &=& \frac{3}{2\sqrt{91}}\, \ell \; \mbox{ with }\ell=1.43\, A^o \mbox{ (C-C bond length)} \\
   \bfx^{(1)} &=& (r,0,0), \mbox{ where }r=\frac{ \sqrt{3} \sqrt{91} }{2\pi} \, \ell \mbox{ (nanotube radius)} \\
   \bfx^{(2)} &=& (\bfR_{\theta_0/3}|\mbox{$\frac{\tau}{3}$}\bfe) \, \bfx^{(1)}.
\end{eqnarray*}
\vspace*{1mm}

\noindent
{\bf Example 2: Tobacco mosaic virus (TMV).} This is a basic example of a filamentous virus, built from a single protein. The protein molecules are arranged in a low-pitch helix, with 16$\frac13$ proteins per turn and with adjacent turns in contact. The parameter values in \eqref{He}, \eqref{H02} are 
\begin{eqnarray*}
    M &=& 1284 \; \mbox{(number of atoms in the protein)} \\
    n &=& 1 \; \mbox{(single helix)} \\
    \theta_0 &=& \frac{2\pi}{16\frac{1}{3}} = \frac{3}{49} \, 2\pi \\
    \tau_0 &=& 1.402 \, A^o.
\end{eqnarray*}
The values for $\theta_0$ and $\tau_0$, taken from \cite{Ge}, correspond to the low-calcium state. 
\\[2mm]
The groups in \eqref{H02} are not some ad hoc ansatz. It can be shown that they are the {\it lattice subgroups} of the helical group \eqref{hel}, whereas the Bravais lattices are the lattice subgroups of the translation group \eqref{trans}.\footnote{In group theory, a subgroup is called a lattice subgroup if it is discrete, i.e. has no accumulation points, and there exists a set of finite volume (Haar measure) whose orbit under the subgroup gives the whole group. To derive the representation \eqref{H0}, one shows and uses that each such subgroup must be the image of a lattice subgroup of the two-dimensional translation group $\R^2$ under the map $h \, : \, \R^2\to \calH_{\bfe}$ defined by $h(\theta,\tau)=(\bfR_{\theta\,\mbox{\scriptsize mod}\, 2\pi}|\tau\bfe)$.
} %end footnote

We now introduce a purely group-theoretical notion of reciprocal lattice. We first state this notion in abstract mathematical language, then show how it reduces to familiar concepts in the crystalline and helical case.
\begin{definition} {\bf (Reciprocal lattice group)} Let $H_0$ be a lattice subgroup of the helical group (see \eqref{H02}), or more generally any lattice subgroup of a locally compact abelian group $H$. Recall from \eqref{dual} the dual group of $H$, $H' = \{\chi\, : \, H \to \C\backslash\{0\}\, : \, \chi \mbox{ is a character of }H\}$. The {\emph reciprocal lattice group} of $H_0$ with respect to $H$ is the set of those characters of $H$ which are equal to 1 on $H_0$, i.e.
$$
   H'_0 = \{\chi \in H'\, : \, \chi(h)=1 \mbox{ for all }h\in H_0\}.
$$
\end{definition}
Thus mathematically, the reciprocal lattice group $H'_0$ is a subgroup of the dual group $H'$.
Physically, the dual group consists of certain scalar waves parametrized by wavevectors or wavenumbers (see the examples below), and the reciprocal lattice group consists of resonant waves, parametrized by a subset of wavevector or wavenumber space which we call the {\it reciprocal lattice}.

In group theory, $H'_0$ is a well known object, called the annihilator of $H_0$
(see e.g. \cite{HewittRoss}) or the orthogonal group of $H_0$ (see e.g. \cite{Reiter}; this terminology views the equation $\chi(h)=1$ as analogous to the vanishing of a proper inner product between elements $\chi$ and $h$ of some vector space) with respect to $H$.
\\[2mm]
{\bf Example 1: Reciprocal lattice of crystals} \\[1mm]
a) {\it Mathematical description.} Let $H$ be the translation group $\calT=\{(\bfI|\bfa)\, : \, \bfa\in\R^d\}$, and let $\calL$ be a Bravais lattice, i.e. $\calL=A\,\Z^d$, $A$ an invertible $d\times d$ matrix. The characters of $H$ as functions of $\bfa\in\R^d$ are known to be the scalar plane waves
$$
    \chi_{\bfk}(\bfa) = e^{i\bfk\cdot\bfa}, \;\;\bfk\in\R^d.
$$
Hence dual group $H'$ and reciprocal lattice group $H'_0$ are
\begin{eqnarray*}
   H' &=& \{ \chi_{\bfk} \, : \, \bfk\in\R^d\}, \\
   H'_0 &=& \{ \chi_{\bfk} \, : \, \bfk\in\R^d, \, \bfk\cdot\bfa\in 2\pi\Z \mbox{ for all }\bfa\in A\,\Z^d\}.
\end{eqnarray*}
b) {\it Physical description.} The characters $\chi_{\bfk}\in H'$ are parametrized by the wavevectors $\bfk\in\R^d$. The wavevector space which parametrizes the group $H'$ is
$$
   \calH' = \R^d
$$
and the set of wavevectors which parametrizes the reciprocal lattice $H'_0$ is
$$
   \calH'_0 = \{\bfk\in\R^d \, : \, \bfk\cdot\bfa\in 2\pi\Z 
   \mbox{ for all }\bfa\in A\,\Z^d\},
$$
recovering the standard definition \eqref{L'} of the reciprocal lattice. 
\\[2mm]
{\bf Example 2: Reciprocal lattice of helical structures.} \\[1mm]
a) {\it Mathematical description.} Let $H$ be the helical group
$\{(\bfR_\theta|\tau\bfe)\, : \, \theta\in[0,2\pi),\,\tau\in\R\}$ (see \eqref{par}), and let $H_0$ be a lattice subgroup, i.e. a group of form \eqref{H02}. The characters of $H$ as a function of $\theta$ and $\tau$ are known to be the cylindrical waves
\begin{equation} \label{helchar}
  \chi_{\alpha,\beta}(\theta,\tau) = e^{i(\alpha\theta + \beta\tau)}, \;\; (\alpha,\beta)\in \Z\times\R.
\end{equation}
Hence dual group $H'$ and reciprocal lattice group $H'_0$ are 
\begin{eqnarray}
     H' &=& \{\chi_{\alpha,\beta}\, : \, (\alpha,\beta)\in\Z\times\R\}, \label{H'} \\
     H'_0 &=& \{\chi_{\alpha,\beta}\, : \, (\alpha,\beta)\in\Z\times\R, \, 
        \begin{pmatrix} \alpha \\ \beta \end{pmatrix} \cdot
        \begin{pmatrix} \theta \\ \tau \end{pmatrix} \in 2\pi\Z \mbox{ for all }
        \begin{pmatrix} \theta \\ \tau \end{pmatrix} \in A\,\Z_n\times\Z\,\mbox{mod}\,
        \begin{pmatrix} 2\pi \\ 0 \end{pmatrix} \}. \label{H0'}
\end{eqnarray}
b) {\it Physical description.} The characters \eqref{helchar} are parametrized by an angular wavenumber $\alpha\in\Z$ and an axial wavenumber $\beta\in\R$, yielding the following parameter space for the dual group $H'$: 
\begin{equation} \label{H'coords}
    \calH' = \{(\alpha,\beta)\, : \, \alpha\in\Z, \, \beta\in\R\} = \Z\times\R.
\end{equation}
The reciprocal lattice is
\begin{equation} \label{H0'coords}
   \calH'_0 = \{(\alpha,\beta)\in\Z\times\R\, : \, 
   \begin{pmatrix} \alpha \\ \beta \end{pmatrix} \cdot
        \begin{pmatrix} \theta \\ \tau \end{pmatrix} \in 2\pi\Z \mbox{ for all }
        \begin{pmatrix} \theta \\ \tau \end{pmatrix} \in A\,\Z_n\times\Z\,\mbox{mod}\,
        \begin{pmatrix} 2\pi \\ 0 \end{pmatrix} \}.      
\end{equation} 
Using the explicit form \eqref{H0el} of $\calH_0$, it follows that $(\alpha,\beta)$ belongs to the reciprocal lattice if and only if 
\begin{equation} \label{resona}
   \alpha \frac{2\pi}{n} \in 2\pi\Z, \;\;\; \alpha\theta_0 + \beta\tau_0 \in 2\pi\Z.
\end{equation}
The solutions $(\alpha,\beta)$ to this condition are easily computed and one obtains, analogously to \eqref{L'explicit}, 
\begin{equation} \label{H0'expl}
   \calH'_0 = 2\pi A^{-T}\Z\times\Z = 
   \begin{pmatrix} n & 0 \\ -\frac{n\theta_0}{\tau_0} & \frac{2\pi}{\tau_0}\end{pmatrix}
   \Z\times\Z = \{ i \begin{pmatrix} n \\ -\frac{n\theta_0}{\tau_0} \end{pmatrix} 
   + j \begin{pmatrix} 0 \\ \frac{2\pi}{\tau_0} \end{pmatrix} \, : \, i,j\in\Z\}.
\end{equation}
Formula \eqref{H0'expl} recovers the reciprocal helical lattice associated with a helical structure generated by the group \eqref{H02} as introduced in \cite{KlugEtAl}, and reveals its group-theoretic meaning as a wavenumber space which parametrizes the group \eqref{H0'}. 
See Figure \ref{fig:H0'}.

\section{Diffraction of twisted waves: twisted Von Laue condition} \label{sec:TVL}
We now calculate the outgoing radiation when twisted waves are scattered off helical structures of infinite length. The reciprocal helical lattice will naturally appear, due to its intrinsic group-theoretic meaning derived in the last section which leads to an associated Poisson summation formula on the helical group. 

Let $\bfE_0(\bfx;\alpha,\beta,\gamma)$ be a twisted wave \eqref{TW} with parameters $(\alpha,\beta,\gamma)\in\Z\times\R\times(0,\infty)$. By eq. \eqref{out}--\eqref{k''}, the outgoing field scattered from a structure with electron density $\rho$ is, assuming $\bfx_c=0$,
\begin{equation} \label{out1}
   \bfE_{out}(\bfx, t) = 
    - c_{e\ell} \frac{e^{i(\bfk(\bfx)\cdot \bfx - \omega t)}}{|\bfx|} 
   \left( \bfI - \frac{\bfk'(\bfx)}{|\bfk'(\bfx)|} \otimes
   \frac{\bfk'(\bfx)}{|\bfk'(\bfx)|}
   \right) \bff_{\bfE_0 } (\bfk'(\bfx) )
\end{equation}
with outgoing wavevector $\bfk'$ and structure factor $\bff_{\bfE_0}$ given by 
\begin{equation} \label{out2}
   \bfk'(\bfx) = \frac{\omega}{c} \frac{\bfx}{|\bfx|}, \hspace*{1cm}  
   \bff_{\bfE_0}(\bfk'(\bfx)) = \int_{\R^3} \bfE_0(\bfy; \alpha,\beta,\gamma )
   \rho(\bfy)e^{-i\bfk'(\bfx)\cdot\bfy} d\bfy.
\end{equation}
The difficulty is to evaluate the structure factor and unearth the fact that it has a sharp peak structure, which is far from obvious from eq. \eqref{out2}. 

We will proceed in 5 steps: (1) transform to cylindrical coordinates; (2) eliminate the ensuing phase nonlinearity of the plane-wave factor inside the integral \eqref{out2} by expanding this plane wave into cylindrical waves, which mathematically corresponds to a Fourier series expansion in the angle variable (this step is not necessary in case of an axial detector); (3) treat the integration over $\varphi'\in S^1$ and $z'\in\R$ jointly rather than separately, because these variables are ``intertwined'' in the electron density of a helical structure, and use Fourier calculus on the group $S^1\times\R$ to essentially reduce the integral to the Fourier transform of an infinite sum of delta functions on $S^1\times\R$; (4) apply the Poisson summation formula on the abelian group $S^1\times\R$, thereby obtaining a twisted analogue of the Von Laue condition in case of an axial detector; (5) eliminate the expansion of Step 2 in case of a non-axial detector. 
\\[2mm]
{\bf Step 1} (Cylindrical coordinates). Let $(r,\varphi,z)$ and $(r',\varphi',z')$ be cylindrical coordinates for $\bfx$ respectively $\bfy$, i.e. $\bfx$ is given by \eqref{cyl} and $y_1=r'\cos\varphi'$, $y_2=r'\sin\varphi'$, $y_3=z'$. Next, let $(R,\Phi,Z)$ be the cylindrical components of the outgoing wavevector field $\bfk'(r,\varphi,z)$, i.e. $\bfk' = R\bfe_r + \Phi \bfe_\varphi + Z \bfe_z$, with the cylindrical unit vectors of eq. \eqref{erephiez}. It follows that 
\beq \label{k'cyl}
   R(r,\varphi ,z) = \frac{\omega}{c} \frac{r}{\sqrt{r^2+z^2}}, \;\; 
   \Phi (r,\varphi, z)=0, \;\; Z(r,\varphi, z) = \frac{\omega}{c} \frac{z}{\sqrt{r^2+z^2}}.
\eeq
With the help of the rotations $\bfR_\varphi$ and $\bfR_{\varphi'}$ (see \eqref{par}), 
we have 
$$
    \bfk'(r,\varphi,z) = \bfR_\varphi \begin{pmatrix} R \\ 0 \\ Z\end{pmatrix}, 
    \hspace*{1cm}
    \bfy = \bfR_{\varphi'} \begin{pmatrix} r' \\ 0 \\ z'\end{pmatrix}.
$$
Consequently the phase of the plane-wave factor in the integral \eqref{out2} is
$$
     \bfk'\cdot\bfy = R \, r' \, \cos(\varphi-\varphi') + Z \, z'.
$$
Substitution into the representation \eqref{TW'} of twisted waves in which the rotation matrix has been diagonalized gives
\begin{align} \label{out3}
 & \bff_{\bfE_0}(\bfk'(r,\varphi,z)) = \bfN(\bfn) \int_0^\infty \int_{S^1\times\R} 
   \rho(r',\varphi',z')\, e^{i(\alpha\varphi' + \beta z')}
   \begin{pmatrix} 
    e^{i\varphi'} & & \\
    & e^{-i\varphi'} & \\
    & & 1 \end{pmatrix}
    \begin{pmatrix} J_{\alpha+1}(\gamma r') \\ J_{\alpha-1}(\gamma r') \\ 
    J_\alpha(\gamma r') \end{pmatrix}   \nonumber \\
 & \hspace*{6.3cm} \cdot e^{-i(R \, r' \, \cos(\varphi-\varphi') + Z\, z')}
    r' dr' d\varphi' dz', 
\end{align}
where here and below, by $\int_{S^1\times\R}d\varphi'dz'$ we mean 
$\int_0^{2\pi}\int_\R d\varphi'dz'$. 
\\[2mm]
{\bf Step 2} (Eliminate the phase nonlinearity). We expand the plane wave factor in \eqref{out3} into cylindrial waves, corresponding to a Fourier series expansion in the angle. The Fourier expansion of the $2\pi$-periodic function $e^{-iA\cos\theta}$ is the Jacobi-Anger expansion 
\beq \label{bes_series}
   e^{-iA\cos\theta} = \sum_{\nu=-\infty}^\infty (-i)^\nu J_\nu(A)e^{i\nu\theta}
   = \sum_{\nu=-\infty}^\infty J_\nu(A) e^{i\nu(\theta-\pi/2)}
\eeq 
where the $J_\nu$ are Bessel functions. It follows that
\begin{align} \label{out5} 
 & \bff_{\bfE_0}(\bfk' (r,\varphi ,z) ) = \bfN(\bfn ) 
  \sum_{\nu=-\infty}^\infty e^{i\nu(\varphi-\pi/2)} \int_0^\infty J_\nu (Rr') \\ 
 & \cdot \Bigl[ \int_{S^1\times\R} \rho(r',\varphi', z')
   \begin{pmatrix} 
    e^{i[(\alpha \pl 1-\nu)\varphi' + (\beta-Z)z']}
              \textcolor{white}{blabla} \textcolor{white}{bla} \\ 
    \textcolor{white}{blabla} e^{i[(\alpha-1-\nu)\varphi'+(\beta-Z)z']} 
              \textcolor{white}{bla} \\ 
    \textcolor{white}{blabla} \textcolor{white}{blabla} 
              e^{i[(\alpha-\nu)\varphi' + (\beta-Z)z']} 
    \end{pmatrix} d\varphi' dz' \Bigr]
    \begin{pmatrix} J_{\alpha+1}(\gamma r') \\ J_{\alpha-1}(\gamma r') \\ 
    J_\alpha (\gamma r') \end{pmatrix}
    r' dr'. \nonumber 
\end{align}
In the special case of an axial detector, i.e. $(r,\varphi,z)=(0,0,z)$, the cylindrical components of the outgoing wavevector $\bfk'(r,\varphi,z)$ are
\beq \label{axialk'}
  R=0, \;\; \Phi = 0, \;\; Z = \frac{\omega}{c}\, \mbox{sign}(z).
\eeq
Thus the phase nonlinearity is absent and we obtain directly from \eqref{out3} that 
\beq \label{out5'} 
  \bff_{\bfE_0}(\bfk' (0, 0, z) ) = \bfN(\bfn ) 
   \int_0^\infty \Bigl[ \int_{S^1\times\R} \rho(r',\varphi', z')
   \begin{pmatrix}
    e^{i[(\alpha \pl 1)\varphi' + (\beta-Z)z']} \textcolor{white}{blabla} \textcolor{white}{bla} \\ 
    \textcolor{white}{blabla} e^{i[(\alpha-1)\varphi'+(\beta-Z)z']} \textcolor{white}{bla} \\ 
    \textcolor{white}{blabla} 
\textcolor{white}{blabla}  e^{i[\alpha\varphi' + (\beta-Z)z']} \end{pmatrix} d\varphi' dz' \Bigr]
    \begin{pmatrix} J_{\alpha+1}(\gamma r') \\ J_{\alpha-1}(\gamma r') \\ 
    J_\alpha (\gamma r') \end{pmatrix}
    r' dr'. \nonumber 
\eeq
Note that in the case \eqref{axialk'}, the sum \eqref{out5} indeed reduces to \eqref{out5'} as it should, because of the property of Bessel functions that 
\begin{equation} \label{besselorth}
     J_\nu(0)=1 \mbox{ for }\nu=0\mbox{ and }0 \mbox{ otherwise}. 
\end{equation}
{\bf Step 3} (Exploit helical symmetry and Fourier calculus on $S^1\times\R$). Consider now a helical structure, i.e. a structure generated by any discrete helical group $H_0$ (see \eqref{H02}). The electron density $\rho$ will be $H_0$-periodic, that is to say
\beq \label{H0per}
   \rho(h^{-1}\bfy) = \rho(\bfy) \mbox{ for all }\bfy\in\R^3\mbox{ and all }h\in H_0,
\eeq
and rapidly decaying in the direction perpendicular to the helical axis.
Typical examples of $H_0$-periodic densities are depicted in Figures 6 and
7. 
By choosing a suitable partition of unity, such a $\rho$ can be written as a sum of rotated and translated copies of a localized, rapidly decaying function $\psi$,  
\beq \label{rho1}
   \rho(\bfy) = \sum_{h\in H_0} \psi(h^{-1}\bfy).
\eeq
In cylindrical coodinates, $H_0$-periodicity means that
$$
   \rho(r',\varphi',z') = \rho(r', \varphi'-\theta\,\mbox{mod}\, 2\pi, z'-\tau)
   \mbox{ for all } \begin{pmatrix}\theta \\ \tau \end{pmatrix}\in \calH_0, \; 
   \calH_0 = A\, \Z_n\times\Z \, \mbox{mod} \, \begin{pmatrix} 2\pi \\ 0 \end{pmatrix},
$$
and the representation \eqref{rho1} means that
\beq \label{rho1'}
  \rho(r',\varphi',z') = \sum_{\bfa\in\calH_0} \psi(r',\varphi'-a_1\,\mbox{mod}\, 2\pi, z' - a_2).
\eeq
The function $\psi$ can for instance taken to be the restriction of $\rho$ to the unit cell $\calU$ of the structure, 
\beq \label{rhoU}
  \psi = \rho_\calU = \begin{cases} \rho & \mbox{in }\calU \\
                             0    & \mbox{outside }\calU \end{cases}, 
  \hspace*{1cm}
  \calU = \{(r',\varphi',z')\, : \, (\varphi',z') \in A\, [0,1]^2, \, r'\in(0,\infty)\},
\eeq
but other constructions with a smooth $\psi$ make sense too. 

The decomposition \eqref{rho1} of $\rho$ can be fruitfully re-written as a convolution of $\psi$ with an infinite sum of delta functions, 
$$
  \rho = \psi \, *_{_{S^1\times\R}} \delta_{\calH_0}, \;\;\; \delta_{\calH_0} = \sum_{\bfa\in\calH_0} \delta_\bfa, 
$$
where the convolution on $S^1\times\R$ is defined as
\begin{equation} \label{convo}
  (f \, *_{_{S^1\times\R}} g)(r,\varphi,z) = \int_{S^1\times\R} f(r,\varphi-\varphi'\,\mbox{mod}\, 2\pi, z-z') \, g(r, \varphi', z')\, d\varphi'dz'. 
\end{equation}
In the sequel we drop the subscript from the convolution sign. 
We now use Fourier calculus on $S^1\times\R$. The Fourier transform $\FSonetimesR$ with respect to the angle and axis variables is a function on the dual group 
$(S^1\times\R)'\stackrel{\sim}{=}\Z\times\R$, defined as 
\beq \label{Fdef} 
  (\FSonetimesR f)(r,\Phi,Z) = \frac{1}{2\pi} \int_{S^1\times\R} f(r,\varphi,z) \, e^{- i (\Phi \, \varphi + Z \, z)} d\varphi\, dz. 
\eeq
The square bracket in \eqref{out5} has the form of such a Fourier transform on $S^1\times\R$: 
$$
  \Bigl[ \;\; \Bigr] = 2\pi 
  \begin{pmatrix}
     \FSonetimesR(\psi *  \delta_{\calH_0})(r',\nu -(\alpha +1), Z-\beta ) 
     \hspace*{5mm} 0 \hspace*{5mm} 0 \\
     0 \hspace*{5mm} \FSonetimesR(\psi *  \delta_{\calH_0})(r',\nu -(\alpha -1), Z-\beta ) \hspace*{5mm} 0 \\
     0 \hspace*{10mm} 0 \hspace*{10mm} \FSonetimesR(\psi *  \delta_{\calH_0})(r',\nu -\alpha, Z-\beta )
  \end{pmatrix}.
$$
{\bf Step 4} (Poisson summation formula on $S^1\times\R$). We now use the (trivial) convolution rule 
on $S^1\times\R$, 
\begin{equation} \label{convol}
 \FSonetimesR(f*g) = 2\pi \, \FSonetimesR f \, \cdot \,  \FSonetimesR g, 
\end{equation}
and the following nontrivial result from Fourier analysis on abelian groups which makes the reciprocal helical lattice appear:
\begin{lemma} {\bf (Poisson summation formula on $S^1\times\R$).}  Let $H_0$ be any discrete helical group (see eq. \eqref{H02}), and let $H'_0$ be the reciprocal lattice. The corresponding parametrizations $\calH_0\subset S^1\times\R$,  \eqref{calH0}, and $\calH'_0\subset\Z\times\R$, \eqref{H0'expl}, satisfy 
\beq \label{groupPoisson}
   \FSonetimesR \delta_{\calH_0} = \frac{2\pi}{|\det A|} \, \delta_{\calH'_0}.
\eeq
\end{lemma}
This identity is a special case of the general Poisson formula on locally compact abelian
groups going back to A.~Weil \cite{Weil}, see e.g. \cite{Reiter}\footnote{In this context, the formula is stated and derived up to an overall multiplicative constant.}. A more elementary derivation of \eqref{groupPoisson} is to first consider the case $\theta=0$, where the result follows by combining the usual Poisson formula on $\R$ with the following, elementary to check, Poisson formula on $S^1$: if $\calL=\{\frac{2\pi j}{n} \, : \, j=0,..,n-1\}$, then 
$$
  \delta_{\calL}(\varphi) = \sum_{j=0}^{n-1} \delta_{\frac{2\pi j}{n}}(\varphi), \; \varphi\in[0,2\pi), \hspace*{1cm} 
  \widehat{\delta_{\calL}}(\nu) = \frac{n}{2\pi} \delta_{n\Z}(\nu) = \frac{n}{2\pi}\sum_{a\in n\Z} \delta_a(\nu), \; \nu\in\Z.
$$
Here $\hat{f}(\nu)$ denotes the Fourier coefficient $(2\pi)^{-1}\int_0^{2\pi}e^{-i\nu\varphi}f(\varphi)\, d\varphi$. 
Note that the delta functions in the left sum are Dirac deltas, whereas the delta functions in the right sum are Kronecker deltas.
The general result \eqref{groupPoisson} now follows from a suitable change of variables.
\\[2mm]  
Eqs.~\eqref{convol}, \eqref{groupPoisson} yield 
\begin{equation} \label{Fdens}
  \FSonetimesR(\psi * \delta_{\calH_0}) = \frac{(2\pi)^2}{|\det A|} (\FSonetimesR\psi) \cdot \delta_{\calH'_0}
\end{equation}
and therefore 
\begin{align} \label{out6} 
 & \bff_{\bfE_0}(\bfk' (r,\varphi ,z) ) = \frac{(2\pi)^3}{|\det A|} \, \bfN(\bfn ) 
  \sum_{\nu=-\infty}^\infty e^{i\nu(\varphi -\pi/2)} D_{\alpha,\beta,\gamma,\nu}(R,Z) 
  \begin{pmatrix} \delta_{\calH'_0}(\nu -(\alpha \pl 1),Z-\beta) \\ 
    \delta_{\calH'_0}(\nu -(\alpha \pl 1),Z-\beta) \\ 
    \delta_{\calH'_0}(\nu -\alpha, Z-\beta) \end{pmatrix}, \\ 
 & \mbox{ with }
   D_{\alpha,\beta,\gamma,\nu}(R,Z) = \int_0^\infty \! J_\nu (Rr') 
   \begin{pmatrix} 
    \FSonetimesR\psi(\nu -(\alpha +1),Z-\beta) J_{\alpha +1}(\gamma r')
              \hspace*{5mm} 0 \hspace*{5mm} 0 \\ 
    0 \hspace*{5mm} 
    \FSonetimesR\psi(\nu -(\alpha-1),Z-\beta) J_{\alpha -1}(\gamma r') \hspace*{5mm} 0 \\ 
    0 \hspace*{12mm} 0 \hspace*{12mm} 
    \FSonetimesR\psi (\nu -\alpha ,Z-\beta ) J_\alpha(\gamma r') 
    \end{pmatrix} r' dr' . \nonumber     
\end{align}
For an axial detector \eqref{axialk'}, the sum over $\nu$ reduces to the contribution from the single term $\nu=0$, thanks to \eqref{besselorth}. Moreover for $\nu=0$ the first Bessel factor, $J_\nu(Rr')$, in the matrix-valued integral in \eqref{out6} equals $1$. Hence the matrix components of the integral involve just one remaining Bessel factor, and can thus be interpreted as a {\it Hankel transform}. Recall that for any $\alpha\in\Z$, the Hankel transform of order $\alpha$ maps scalar functions of a radial variable belonging to the interval $(0,\infty)$ to scalar functions on $(0,\infty)$, and is defined as
\beq \label{Hankel}
   (H_\alpha f)(\gamma) = \int_0^\infty \! f(r') \, J_\alpha(\gamma r') \, r' dr' \;\;\; (\gamma > 0).
\eeq
It follows that 
\begin{align} \label{out7}
 & \bff_{\bfE_0}(\bfk' (0,0,z) ) = \frac{(2\pi)^3}{|\det A|} \, \bfN(\bfn ) D_{\alpha,\beta,\gamma}(Z) 
   \begin{pmatrix} \delta_{\calH'_0}(-(\alpha \pl 1),Z-\beta) \\ 
    \delta_{\calH'_0}(-(\alpha \pl 1),Z-\beta) \\ 
    \delta_{\calH'_0}(-\alpha, Z-\beta) \end{pmatrix}, \\
 & \mbox{with }
   D_{\alpha,\beta,\gamma}(Z) = 
   \begin{pmatrix} 
    H_{\alpha+1}\FSonetimesR\psi(\gamma, -(\alpha +1), Z-\beta)
              \hspace*{5mm} 0 \hspace*{5mm} 0 \\ 
    0 \hspace*{5mm} 
    H_{\alpha-1}\FSonetimesR\psi(\gamma,-(\alpha-1),Z-\beta) \hspace*{5mm} 0 \\ 
    0 \hspace*{12mm} 0 \hspace*{12mm} 
    H_{\alpha}\FSonetimesR\psi (\gamma, -\alpha ,Z-\beta ) 
    \end{pmatrix} .    
\end{align}
For the purpose of calculating the outgoing intensity, it is useful to write the delta function $\delta_{\calH'_0}$ as a sum over reciprocal lattice vectors, $\delta_{\calH'_0}=\sum_{\bfa\in\calH'_0}\delta_{\bfa'}$, and note that
$$
   \delta_{\bfa'}(-(\alpha\pm 1),Z-\beta) = \delta_{\bfa' \pm \mbox{\tiny $\begin{pmatrix} 1 \\ 0 \end{pmatrix}$}} (-\alpha, Z-\beta). 
$$
This yields the following alternative expression for the structure factor:
\beq \label{out7'}
   \bff_{\bfE_0}(\bfk' (0,0,z) ) = \frac{(2\pi)^3}{|\det A|} \, \bfN(\bfn ) 
    \sum_{\bfa'\in\calH'_0} \Bigl( H_{-a'_1}\FSonetimesR\psi\Bigr)(\gamma,\bfa')
    \begin{pmatrix} 
    \delta_{\bfa' + \mbox{\tiny $\begin{pmatrix} 1 \\ 0 \end{pmatrix}$}} \\ 
    \delta_{\bfa' - \mbox{\tiny $\begin{pmatrix} 1 \\ 0 \end{pmatrix}$}} \\ 
    \delta_{\bfa'} 
    \end{pmatrix} (-\alpha, Z-\beta). 
\eeq
Note that the delta functions in \eqref{out7'} are centered on shifted copies of the reciprocal lattice. Also, we claim that the Fourier-Hankel transform of $\psi$ which appears in \eqref{out7'} equals that of the electron density \eqref{rhoU} in the unit cell, i.e.
\beq \label{FrhoU}
   H_{-a'_1}\FSonetimesR\psi(\gamma,\bfa') 
   = H_{-a'_1}\FSonetimesR\rho_\calU(\gamma,\bfa') \;\; \mbox{ for all }\bfa'\in\calH'_0.
\eeq
This is because, for any $\psi$ satisfying \eqref{rho1}, including $\psi=\rho_U$, the Fourier transform $(\FSonetimesR\psi)(r',\bfa')$ is independent of $\psi$ when $\bfa'$ is a reciprocal lattice vector; indeed
\begin{eqnarray*}
  (\FSonetimesR\psi)(R',\bfa') & = & \int_{S^1\times\R} e^{i \bfa' \cdot \mbox{\tiny $\begin{pmatrix}\varphi' \\ z' \end{pmatrix}$}} \psi(r',\varphi',z')d\varphi' dz' \\
   & = & \int_{\calU} e^{- i \bfa' \cdot (\mbox{\tiny $\begin{pmatrix}\varphi' \\ z' \end{pmatrix}$} - \bfa)} \psi(r', \varphi' - a_1 \, \mbox{mod} \, 2\pi, z'-a_2) \, d\varphi' dz' \\
   & = & \int_{\calU} e^{- i \bfa' \cdot \mbox{\tiny $\begin{pmatrix}\varphi' \\ z' \end{pmatrix}$}} \rho(r',\varphi',z') \, d\varphi' dz' \mbox{ for all }\bfa'\in\calH'_0.
\end{eqnarray*}  
Here we have used that the factor $e^{i\bfa'\cdot\bfa}$ in the middle line equals $1$, and have employed \eqref{rho1'}. 

The outgoing electric field can now be read off immediately from \eqref{out1}, \eqref{out2}, \eqref{out7'}, \eqref{FrhoU}. Note in particular that the third field component vanishes, since the projection matrix in \eqref{out1} annihilates the third component of the structure factor when the outgoing wavevector $\bfk'$ points in axial direction $\pm \bfe_3$. See eq.~\eqref{scat1} below.

When taking absolute values to obtain the intensity, it may happen that the two remaining shifted copies $\calH'_0+(1,0)$ and $\calH'_0-(1,0)$ of the reciprocal lattice appearing in \eqref{out7'} overlap, which would lead to interference. These two copies overlap if and only if $(2,0)$ is a reciprocal helical lattice vector. The following result, which is elementary to check, shows that this does not happen except in a degenerate case which we propose to denote {\it flat helical groups}.
\begin{lemma} \label{L:flat} {\bf (Flat helical groups).} The following three statements about a discrete helical group \eqref{H0} are equivalent: \\[1mm]
$\;$ {\rm (1)} $\;$ The vector $(2,0)$ belongs to the reciprocal helical lattice, eq. \eqref{H0'expl}. \\
$\;$ {\rm (2)} $\;$ The parameters of the helical group satisfy $\theta_0=0$ or $\pi$, and $n=1$ or $2$. \\
$\;$ {\rm (3)} $\;$ The structure generated by applying the helical group to any single point lies in a plane.
\end{lemma}
\vspace*{1mm}

We summarize our findings as a theorem. 
\begin{theorem} \label{TVL} {\bf (Twisted Von Laue condition).} Consider a helical structure with electron density $\rho\, : \, \R^3 \to \R$, assumed to be smooth, $H_0$-periodic with respect to some discrete helical group $H_0$ (see eq. \eqref{H0}), and rapidly decaying in the direction perpendicular to the helical axis.  
Assume that the axis is $\bfe=\bfe_3$, and let the incoming electric field be a twisted wave with same axis and parameter vector  $(\alpha,\beta,\gamma)\in\Z\times\R\times(0,\infty)$ 
(see \eqref{TW}--\eqref{N}). Recall that the frequency of this twisted wave is $\omega=c|(0,\gamma,\beta)|$. Then the diffracted electric field \eqref{out1} at any point $(0,0,z)$ on the axis is
\begin{eqnarray} \label{scat1}
  & & \bfE_{out}(0,0,z,t) = -c_{e\ell} \frac{e^{i(\frac{\omega}{c}|z| - \omega t)}}{|z|} \frac{(2\pi)^3}{|\det A|} \sum_{\bfa'\in\calH'_0} 
  (H_{-a'_1}\FSonetimesR \rho_\calU)(\gamma,\bfa') \\
  & & \hspace*{2.5cm} \cdot
   \begin{pmatrix}  
 \mbox{$\frac{n_1+in_2}{2}$} & \mbox{$\frac{n_1-in_2}{2}$} & 0 \\
 \mbox{$\frac{n_2-in_1}{2}$} & \mbox{$\frac{n_2+in_1}{2}$} & 0 \\
  0 & 0 & 0     
   \end{pmatrix}   
   \begin{pmatrix} 
    \delta_{\bfa' + \mbox{\tiny $\begin{pmatrix} 1 \\ 0 \end{pmatrix}$}} \\ 
    \delta_{\bfa' - \mbox{\tiny $\begin{pmatrix} 1 \\ 0 \end{pmatrix}$}} \\ 
       0 
   \end{pmatrix} (-\alpha, \mbox{$\frac{\omega}{c}$} \mbox{\rm sign}\, z -\beta).  \nonumber
\end{eqnarray}
Here $H_\alpha$ is the Hankel transform of order $\alpha$ with respect to the radial variable (see \eqref{Hankel}), $\FSonetimesR$ is the Fourier series/transform with respect to the angular and axial variables (see \eqref{Fdef}), $\rho_\calU$ is the restriction of the electron density $\rho$ to the unit cell (see \eqref{rhoU}), and   
$\calH'_0$ is the reciprocal helical lattice (see \eqref{H0'}). 

Moreover, when $H_0$ is not a flat helical group (see Lemma \ref{L:flat}), the square root of the outgoing intensity is 
\begin{eqnarray} \label{scat2}
   & & \Bigl(I(0,0,z;\, \alpha,\beta,\gamma)\Bigr)^{1/2} = c_0 \frac{(2\pi)^3}{|\det A|}
   \frac{1}{|z|} \sum_{\bfa'\in\calH'_0} \Bigl| (H_{-a'_1}\FSonetimesR\rho_\calU)(\gamma,\bfa')\Bigr|  \\
   & & \hspace*{2.5cm} \cdot 
   \Bigl( \frac{|n_1+in_2|}{\sqrt{2}} 
   \delta_{\bfa' + \mbox{\tiny $\begin{pmatrix} 1 \\ 0 \end{pmatrix}$}} \, + \, 
   \frac{|n_1-in_2|}{\sqrt{2}} 
   \delta_{\bfa' - \mbox{\tiny $\begin{pmatrix} 1 \\ 0 \end{pmatrix}$}} \Bigr)
   (-\alpha, \, \mbox{$\frac{\omega}{c}$}\, \mbox{\rm sign} \, z - \beta), \nonumber
\end{eqnarray}
where $c_0=(\frac{c\eps_0}{2})^{1/2}c_{e\ell}$. In particular, constructive interference occurs if and only if the difference between the angular/axial part $(0,\frac{\omega}{c}\,\mbox{\rm sign}\, z)$ of the outgoing wavevector and the angular/axial parameters $(\alpha,\beta)$ of the incoming twisted wave belongs to the reciprocal helical lattice shifted left or right by precisely one angular wavenumber, $\calH'_0\pm 
\mbox{\tiny $\begin{pmatrix}1 \\ 0 \end{pmatrix}$}$, or equivalently, if and only if
\begin{equation} \label{reson2}
   -\alpha = i \, n \pm 1, \;\;\; \tau_0(\mbox{$\frac{\omega}{c}$}\,\mbox{\rm sign} \, z - \beta) = - i \, n \, \theta_0 + 2\pi j \;\;\mbox{ for some integers }i,j.
\end{equation}
\end{theorem}
Formulae \eqref{scat1}, \eqref{scat2}, \eqref{reson2} are the main result of this paper. They say that the signal of a helical structure under co-axial twisted X-rays, recorded along the axis, consists of sharp peaks with respect to the angular and axial radiation parameters. The peaks are {\it double-peaks} with a distance of precisely two angular wavenumbers, centered at the reciprocal lattice vectors of the helical structure. In particular, at the reciprocal lattice vectors themselves the signal vanishes. The structural parameters $\tau_0$, $\theta_0$ in \eqref{H0}, or equivalently the pitch and the number of subunits per turn, can be immediately read off from the peak locations, as can the order $n$ of any rotational symmetry. 

Moreover the above result makes it in principle possible to determine the electron density $\rho$, i.e. the detailed atomic structure, from intensity measurements in the far field at a specific point on the axis, provided the scalar phase problem associated with the Fourier-Hankel transform $H_\alpha\FSonetimesR$ can be solved. Note that the values of this transform on the reciprocal helical lattice points which appear in \eqref{scat2} completely determine the electron density; see Section \ref{sec:synth} for details.  

Next, we report an important invariance property of the axial signal of a helical structure. 
Suppose the structure is translated by an amount $\Delta z$ along the helical axis, and rotated by an amount $\Delta\varphi$ around the axis. (In other words, we apply an arbitrary element of the continuous helical group \eqref{hel} to the structure.) This modifies the original electron density $\rho_\calU$ to 
\begin{equation}
   {\rho_\calU}'(r,\varphi,z) = \rho_\calU(r,\varphi-\Delta\varphi, z-\Delta z).
\end{equation}
From the definition of the angular/axial Fourier transform, \eqref{Fdef}, it is clear that 
\beq\label{invar}
   \Bigl(\FSonetimesR{\rho_\calU}' \Bigr)(r,\Phi,Z) = e^{-i(\Phi\, \Delta\varphi + Z \, \Delta z)} \Bigl(\FSonetimesR\rho_\calU\Bigr)(r,\Phi,Z).
\eeq
Thus this transform changes just by a phase factor.
The Hankel transform $H_\alpha$ with respect to the radial variable does not interfere with this phase factor, and so it follows that 
\beq\label{invar2}
   \Bigl(H_\alpha\FSonetimesR\rho'_\calU\Bigr)(\gamma,\Phi,Z) = 
   e^{-i(\Phi\, \Delta\varphi + Z \, \Delta z)} 
   \Bigl(H_\alpha\FSonetimesR\rho_\calU\Bigr)(\gamma,\Phi,Z) \;\;\; \mbox{ for all }\gamma,\, \Phi, \, Z.
\eeq
In particular, the outgoing intensity \eqref{scat2}, which only depends on the absolute value of the expression \eqref{invar2}, is {\it invariant} under axial translations and rotations of the structure. This is not a fortuitous accident, but stems from the fact that the design equations, Def.~\eqref{DE}, require the incoming wave to be an eigenfunction of each element of the continuous helical group \eqref{hel}. From this one easily sees that the invariance remains true for helical structures of finite length. 

By comparison, the signal produced by fiber diffraction, i.e. by sending plane wave X-rays towards a helical structure from a perpendicular direction, is only invariant under axial translation but {\it not} under axial rotation; the latter is a well known major problem in the interpretation of fiber diffraction images. 
\\[2mm]
{\bf Step 5} (Arbitrary outgoing direction). Finally, we calculate the outgoing radiation in arbitrary direction. Before doing so, we note the simple group-theoretic reason for why one expects the radiation to be much more complicated when dropping the restriction to axial outgoing wavevectors. Only in this case is the factor $e^{-i\bfk'(\bfx)\cdot\bfy}$ appearing in \eqref{out2} a character of the helical group $\calH_\bfe$, i.e. an element of the dual group $H'$ of the helical group, which -- as explained in Section \ref{sec:TW} -- is the parameter space for twisted waves. But this is necessary in order that the outgoing wavevector just causes a shift of the incoming radiation parameters as in \eqref{scat1}--\eqref{scat2}. 

Indeed, the outgoing radiation in non-axial direction is more complicated. We start from the expression \eqref{out6}, and focus on the first component of the (vector-valued) sum over $\nu$. We have, for any three functions $f$, $g$, $h$ on $\Z$, 
$$
 \sum_{\nu} f(\nu) \, g(\nu - (\alpha \pl 1)) \, h(\alpha \pl 1) \delta_{\calH'_0}(\nu - (\alpha \pl 1), Z-\beta)  = \sum_{\bfa'\in\calH'_0} f(a'_1 + \alpha \pl 1) \, g(a'_1) \, h(\alpha \pl 1) \, \delta_{a'_2}(Z-\beta). 
$$
It follows that  
\begin{eqnarray} \label{out8} 
   & &
  \bff_{\bfE_0}(\bfk' (r,\varphi,z) ) = \frac{(2\pi)^3}{|\det A|} \, \bfN(\bfn )
   \sum_{\bfa'\in\calH'_0}\! \delta_{a'_2}(Z-\beta)
   D_{a'_1,\alpha}(\varphi) \,  
   \int_0^\infty \!\! (\FSonetimesR\psi )(r',\bfa') 
   \begin{pmatrix} 
     J_{a'_1+\alpha + 1}(Rr') J_{\alpha+1}(\gamma r') \\
     J_{a'_1+\alpha - 1}(Rr') J_{\alpha-1}(\gamma r') \\
     J_{a'_1+\alpha}(Rr') J_{\alpha}(\gamma r')
   \end{pmatrix} r'dr' \nonumber  \\
   & & \;\;\; \mbox{ with } \; 
   D_{a'_1,\alpha}(\varphi) = 
   \begin{pmatrix} 
    e^{i(a'_1+\alpha \pl 1)(\varphi-\pi/2)} & 0 & 0 \\ 
    0 & e^{i(a'_1+\alpha-1)(\varphi-\pi/2)} & 0 \\
    0 & 0 & e^{i(a'_1+\alpha)(\varphi-\pi/2)} 
    \end{pmatrix}.
\end{eqnarray}
Denoting the product $f(r',\varphi',z')=J_\nu(Rr')\psi(r',\varphi',z')$ by $f=J_\nu(R\, \cdot\, )\psi$ and using eq. \eqref{rhoU}, this formula for the structure factor can be written more compactly as
\begin{equation} \label{out8'}
   \bff_{\bfE_0}(\bfk' (r,\varphi,z) ) = \frac{(2\pi)^3}{|\det A|} \, \bfN(\bfn ) \sum_{\bfa'\in\calH'_0} \! \delta_{a'_2}(Z-\beta)
   D_{a'_1,\alpha}(\varphi)
   \begin{pmatrix} 
     H_{\alpha+1}\FSonetimesR\Bigl(J_{a'_1+\alpha + 1}(R \, \cdot \, ) \rho_\calU\Bigr) \\
     H_{\alpha-1}\FSonetimesR\Bigl(J_{a'_1+\alpha - 1}(R \, \cdot \, ) \rho_\calU\Bigr) \\
     H_{\alpha}\FSonetimesR\Bigl(J_{a'_1+\alpha}(R \, \cdot \, ) \rho_\calU\Bigr) \\     
   \end{pmatrix}(\gamma, \bfa' ). 
\end{equation}
This is the generalization of formula \eqref{out7'} to an arbitrary outgoing direction. Thus the far field observed from any direction still exhibits a sharp peak structure in the axial wavenumber $\beta$. More precisely, by \eqref{H0'expl}, resonance occurs when 
\begin{equation}
    \tau_0(Z-\beta) = - i\, n\, \theta_0 + 2\pi j \;\;\mbox{ for some integers }i,j.
\end{equation}
But the signal is spread out over angular wavenumbers $\alpha$ that need not be related to the reciprocal lattice. In the limit of the radial component $R$ of the outgoing wavevector tending to zero, the Bessel factors $J_{a'_1+\alpha+\sigma}(R \, \cdot \, )$ ($\sigma=0,1,-1$) in front of the unit cell electron density $\rho_\calU$ converge to the delta functions $\delta_{a'_1+\alpha+\sigma}$, reducing \eqref{out8'} back to the axial formula \eqref{out7'}. 

If the structure factor is {\it averaged} over outgoing wavevecors with fixed angle to the helical axis, sharp peaks appear also with respect to the angular wavenumber. More precisely, the average of the diagonal matrix $D_{a'_1,\alpha}(\varphi)$ over $\varphi$ is a diagonal matrix of delta functions, eg. the upper left entry is $1$ if $a'_1+\alpha+1=0$ and zero otherwise. As a consequence, 
\beq
   \frac{1}{2\pi}\!\int_0^{2\pi}\!\!\!\!\bff_{\bfE_0}(\bfk' (r,\varphi,z) )\, d\varphi = \frac{(2\pi)^3}{|\det A|} \, \bfN(\bfn ) 
    \sum_{\bfa'\in\calH'_0} \Bigl( H_{-a'_1}\FSonetimesR(J_0(R\, \cdot \, )\rho_{\cal U})\Bigr)(\gamma,\bfa')
    \begin{pmatrix} 
    \delta_{\bfa' + \mbox{\tiny $\begin{pmatrix} 1 \\ 0 \end{pmatrix}$}} \\ 
    \delta_{\bfa' - \mbox{\tiny $\begin{pmatrix} 1 \\ 0 \end{pmatrix}$}} \\ 
    \delta_{\bfa'} 
    \end{pmatrix} (-\alpha, Z-\beta). 
\eeq
The only difference as compared to the structure factor \eqref{out7'} in axial direction is the zero$^{th}$ order Bessel factor in front of the unit cell electron density. In particular, by \eqref{H0'expl}, resonance of the $k^{th}$ component of the averaged structure factor occurs when 
\begin{equation} \label{rdj}
   - \alpha = i\, n + \sigma_k, \;\;\; \tau_0(Z-\beta) = - i\, n\, \theta_0 + 2\pi j \;\; \mbox{ for some integers }i,j,
\end{equation}
where $\sigma_k=1$ for $k=1$, $-1$ for $k=2$, and $0$ for $k=3$. 
If this averaged signal can be realized experimentally, the observation direction (which influences the axial component $Z$ of $\bfk'$) can be varied to detect resonance. 
\section{Synthesis of electron density} \label{sec:synth}
As shown in the previous section, subjecting a helical structure to twisted X-rays and recording the intensity of the scattered radiation in axial direction yields the data set 
\begin{equation} \label{data}
   \left\{ |G(\gamma,\alpha,\beta)|^2 \, : \, \gamma\in(0,\infty), \, (\alpha,\beta)\in \calH'_0 
   \right\},
\end{equation}
where $G$ is the Fourier-Hankel transform of the unit cell electron density $\rho_\calU$, 
\beq \label{FH}
    G(\gamma,\alpha,\beta) = \Bigl( H_\alpha 
\FSonetimesR\rho_\calU\Bigr)(\gamma,\alpha,\beta) = \frac{1}{2\pi} 
    \int_0^\infty \iint_\calU e^{-i(\alpha\varphi + \beta z)} J_\alpha(\gamma r)\,
    \rho_U(r,\varphi,z) \, r\, dr\, d\varphi\, dz
\eeq    
and $\calH'_0$ is the reciprocal helical lattice of the structure. Assume that the phases of the Fourier-Hankel coefficients $G$ in \eqref{data} can be retrieved. Then the electron density in the unit cell is recovered by 
\begin{eqnarray} \label{eq:synth}     
   \rho_\calU(r,\varphi,z) &=& \frac{1}{2\pi} \sum_{\mbox{\tiny $\begin{pmatrix}\alpha \\ \beta
   \end{pmatrix}$} \in \calH'_0} \Bigl(H_\alpha G(\, \cdot \, , \alpha,\beta)\Bigr)(r) \, e^{i(\alpha\varphi + \beta z)}  \nonumber \\
    &=& \frac{1}{2\pi} \sum_{\mbox{\tiny $\begin{pmatrix}\alpha \\ \beta
   \end{pmatrix}$} \in \calH'_0} \int_0^\infty \!\! J_\alpha(\gamma r) \, G(\gamma,\alpha,\beta)
   \, \gamma \, d\gamma \,   e^{i(\alpha\varphi + \beta z)}.   
\end{eqnarray}
To show this, one first notes that the Fourier-Hankel transform in \eqref{FH} is a combination of a Fourier series in the angle, a Fourier transform in the axial variable, and a Hankel transform in the radial variable, and is thus invertible. One then exploits formula \eqref{Fdens}
with $\psi=\rho_U$, and uses that the inverse of the Hankel transform \eqref{Hankel} is given by 
\beq \label{Hankelinv}
  \Bigl( H_\alpha^{-1}\tilde{f} \Bigr)(r) = \int_0^\infty \tilde{f}(\gamma) \, J_\alpha(\gamma r) \, \gamma \, d\gamma.
\eeq 
The overall situation is thus the same as in standard X-ray crystallography: the electron density can be reconstructed provided we can solve a scalar phase problem. The only difference is that we have to deal with a 
\\[2mm]
{\bf New phase problem.} {\it Reconstruct a function $\rho_U$ given the absolute value of its Fourier-Hankel transform $G$ defined by \eqref{FH}.}
\\[2mm]
Numerical investigations which will be reported elsewhere \cite{FJJ3} indicate that standard phase retrieval algorithms for the Fourier transform can be adapted without difficulty to the Fourier-Hankel case.
\section{Simulated diffraction pattern of carbon~nanotube and tobacco mosaic virus} \label{sec:numerics}
Here we present simulated diffraction patterns of helical structures subjected to twisted waves. The set-up is as in Figure \ref{fig:setup} of the Introduction, that is to say incoming waves, structure and detector are axially aligned. As pointed out earlier (see \eqref{invar2}), the signal is invariant under axial translations and rotations of the structure. 

We consider two examples, a (6,5)-Carbon nanotube and TMV virus. The parameters are as in Example 1 respectively 2 of Section \ref{sec:reciprocal}, with the TMV atomic positions taken from the Protein Data Bank, PDB ID 3J06 \cite{Ge}. We use the electromagnetic model \eqref{out1}, \eqref{out2}, \eqref{Iout} to calculate the signal from structures of finite length:

\begin{itemize}
\item {\bf C nanotube:} 255 unit cells, corresponding to 510 atoms.
\item {\bf TMV:} 147 proteins, corresponding to 3 helical repeats and 188~748 atoms.
\end{itemize}

\begin{figure}[http!]
\begin{center}
\begin{minipage}{0.65\textwidth} 
\hspace*{1.2mm}\includegraphics[width=0.974\textwidth]{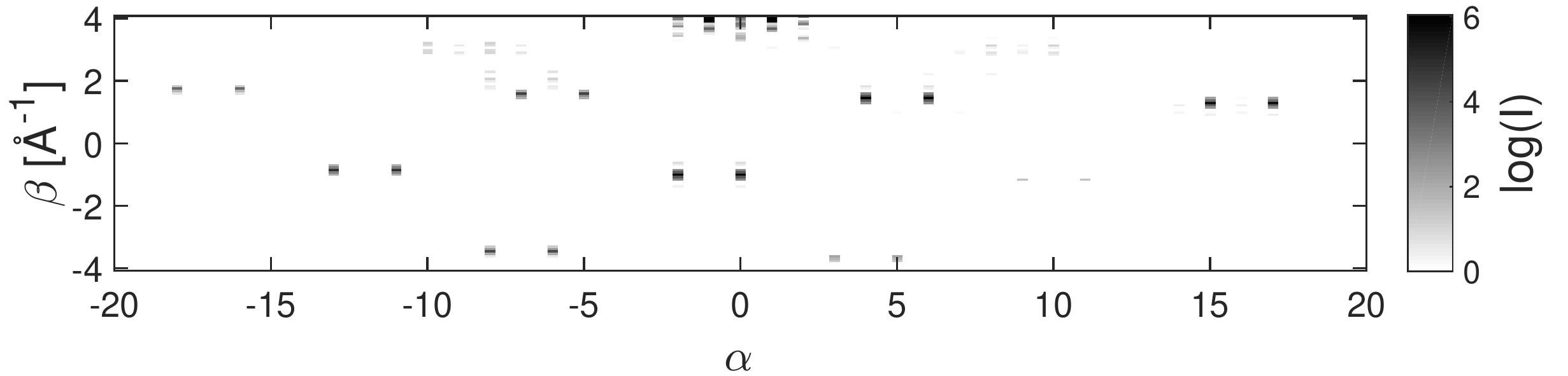} \\
\includegraphics[width=\textwidth]{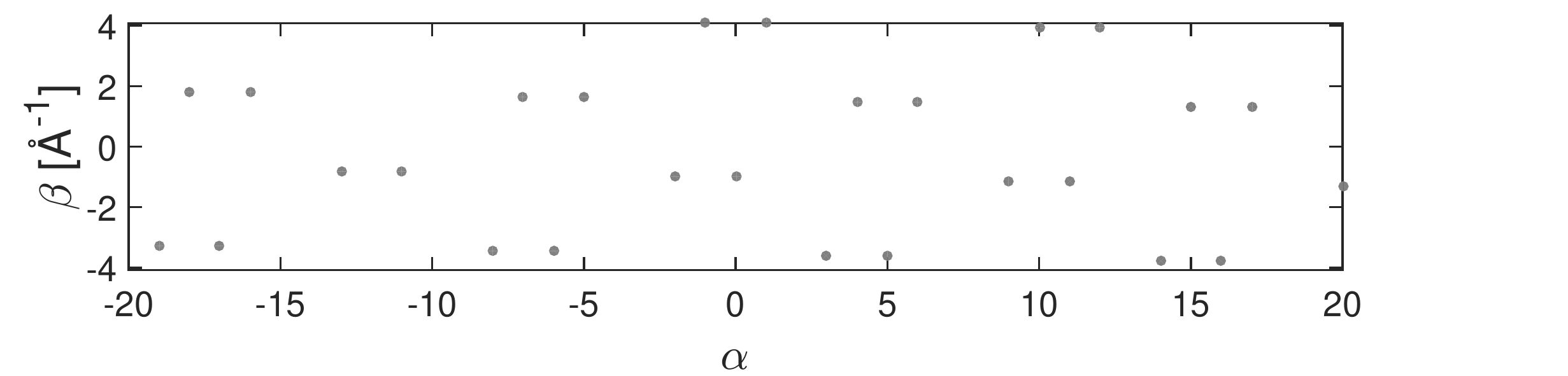}
\end{minipage} \hfill
\begin{minipage}{0.33\textwidth}
\includegraphics[width=\textwidth]{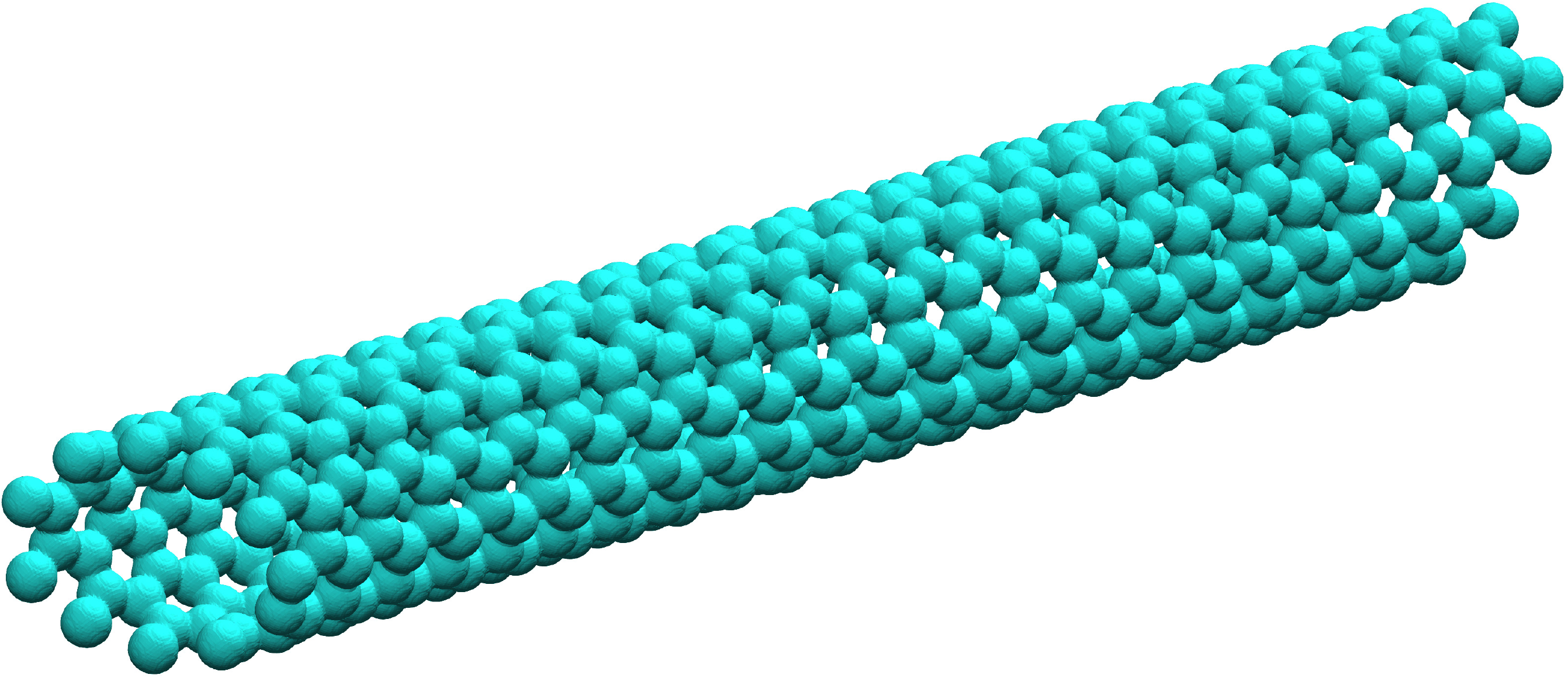} 
\end{minipage} 
\caption{Twisted X-ray pattern of a (6,5)-carbon nanotube with 510 atoms. The electron density of each atom was modelled by a Gaussian. Right: Iso-surface of electron density. Top: Log intensity of diffracted radiation as a function of angular wavenumber $\alpha$ and axial wavenumber $\beta$ of incoming wave. Bottom: Theoretical peak locations, eq. \eqref{scat2}. Incoming twisted X-ray wavelength: $\lambda=1.54$~A$^o$ (Cu $K_\alpha$ line).}
\end{center}
\label{fig:C}
\end{figure}

\begin{figure}[http!] 
\begin{center}
\begin{minipage}{0.65\textwidth} 
\hspace*{1.2mm}\includegraphics[width=0.974\textwidth]{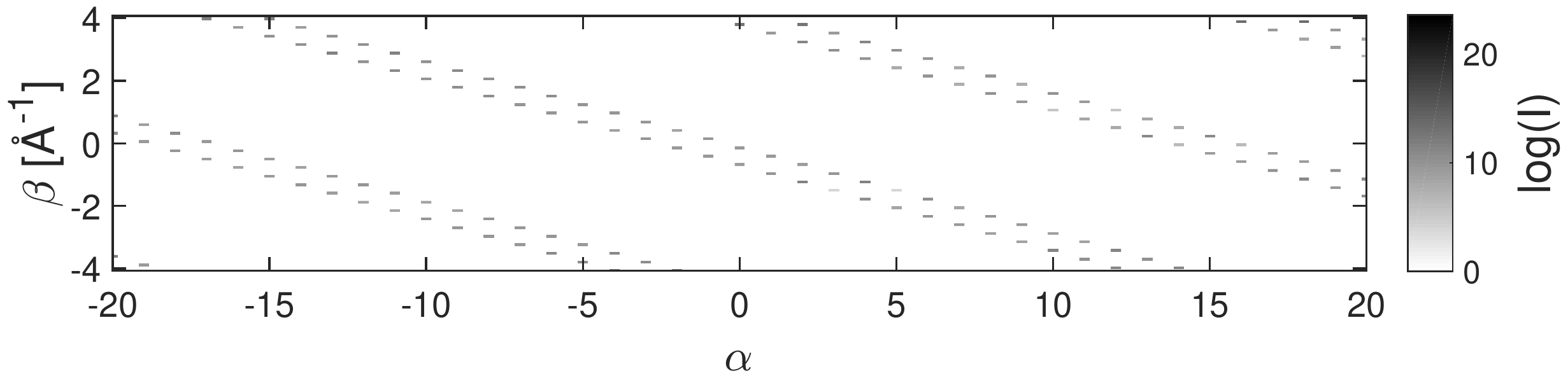} \\
\includegraphics[width=\textwidth]{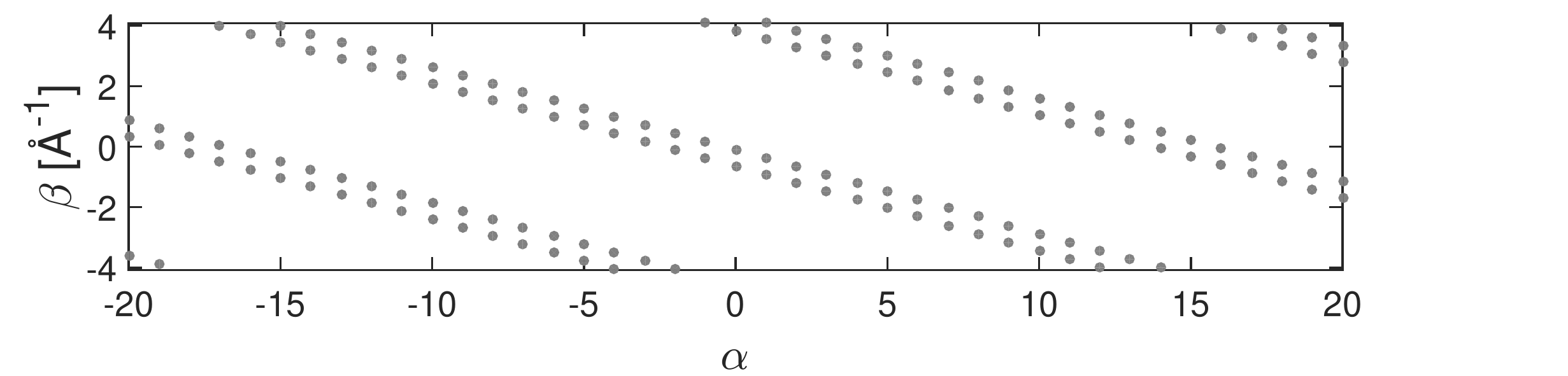} 
\end{minipage} \hfill
\begin{minipage}{0.33\textwidth}
\hspace*{5mm} \includegraphics[width=0.65\textwidth]{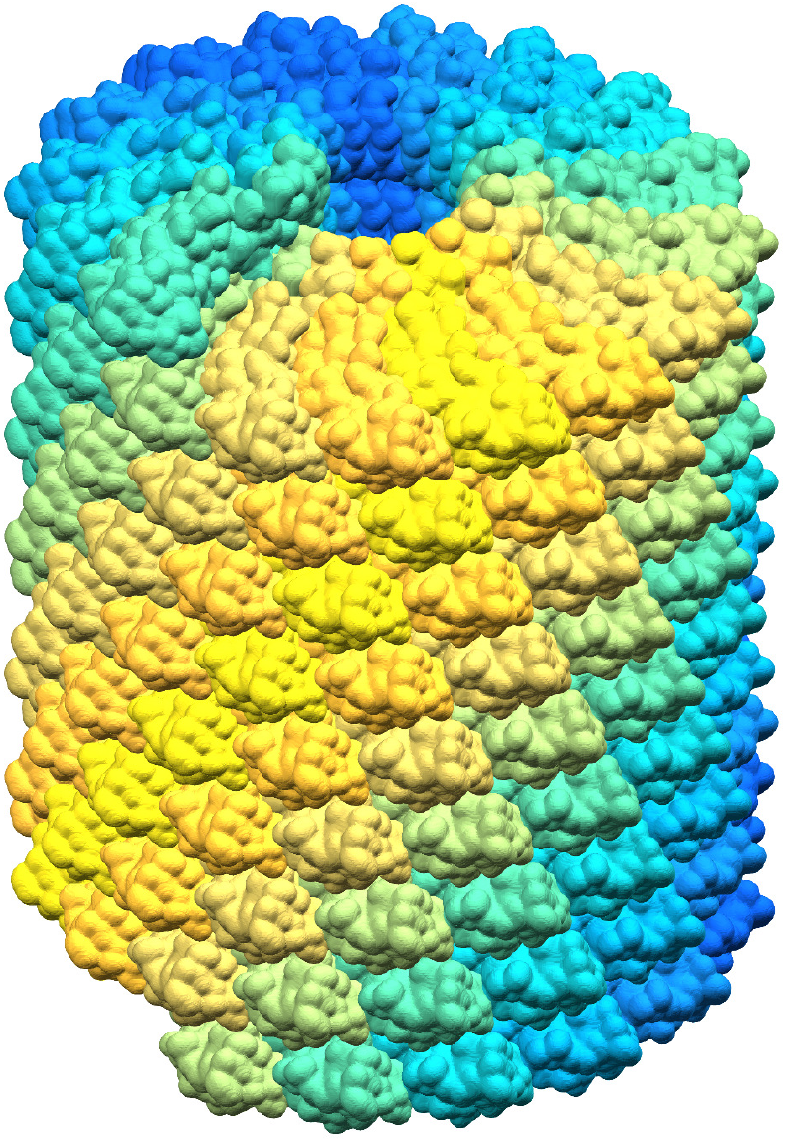} 
\end{minipage} 
\caption{Twisted X-ray pattern of 3 helical repeats of TMV (147 proteins, 188~748 atoms). For simplicity, atomic form factors were ignored and the electron density of each atom was modelled as a delta function. Right: Iso-surface of electron density. Top: Log intensity of diffracted radiation as a function of angular wavenumber $\alpha$ and axial wavenumber $\beta$ of incoming wave. Bottom: Theoretical peak locations, eq. \eqref{scat2}. Incoming twisted X-ray wavelength: $\lambda=1.54$~A$^o$ (Cu $K_\alpha$ line).}
\end{center}
\label{fig:TMV}
\end{figure}

The results in Figures 5 and 6 show that already relatively short pieces of a helical structure exhibit the theoretically predicted patterns. In particular, one sees double-peaks centered at the reciprocal helical lattice points which would be difficult to interpret correctly in terms of scalar models of the incoming wave and the diffraction pattern. Moreover, very high contrast between peaks and background is observed; note that in fiber diffraction the contrast in the direction perpendicular to the fiber is limited for theoretical reasons \cite{CCV52}. 
\section{Fourier representation of twisted waves} \label{sec:Fourier}
We now derive the expansion of twisted waves into plane waves, or mathematically: the Fourier decomposition of twisted waves. This leads to a simple physical interpretation of the polarization vector $\bfn$ and the ``cartesian reduced wavevector'' $\bfk_0 = (0,\gamma,\beta)$ appearing in \eqref{TW}--\eqref{N}, both of which have remained somewhat mysterious up to now. It also suggests a possible route to realizing approximate twisted waves experimentally. 

Consider, e.g. the axial component of a twisted wave, $const \, e^{i(\alpha\varphi + \beta z)}J_\alpha(\gamma r)$. This expression is a scalar cylindrical harmonic, evaluated at the point with cylindrical coordinates $(r,\varphi,z)$. The cartesian coordinates of this point are $x_1=r\cos\varphi$, $x_2=r\sin\varphi$, $x_3=z$. Using the Bessel integral
$$
    J_\alpha(A) = \frac{1}{2\pi} \int_0^{2\pi} \! e^{i(\alpha\varphi' - A \sin \varphi')} d\varphi'
$$
gives
\begin{equation} \label{FourTW1}
   e^{i(\alpha\varphi + \beta z)} J_\alpha(\gamma r) 
   = \frac{1}{2\pi}\int_0^{2\pi} e^{i[\alpha(\varphi'+\varphi)-\gamma r \sin\varphi' + \beta z]}
            d\varphi'  
   = \frac{1}{2\pi}\int_0^{2\pi} e^{i[\alpha\Phi-\gamma r \sin(\Phi-\varphi) + \beta z]}
            d\Phi,
\end{equation}
where we have employed the substitution $\varphi'+\varphi=\Phi$. The phase appearing in the latter integral can be interpreted as a
 cartesian inner product. We have, using the rotation matrices $\bfR_\varphi$ and $\bfR_{\Phi}$ (see \eqref{par}) and the addition formula for the sine, 
\begin{equation} \label{FourTW2}
   - \gamma r \sin(\Phi - \varphi) + \beta z = 
   \begin{pmatrix} - \gamma \sin\Phi \\ \gamma\cos\Phi \\ \beta \end{pmatrix}
   \cdot \begin{pmatrix} r\cos\varphi \\ r \sin\varphi \\ z \end{pmatrix}
   = \bfR_\Phi \bfk_0 \cdot \bfx,
\end{equation}
with $\bfk_0=(0,\gamma,\beta)$. Eqs. \eqref{FourTW1}--\eqref{FourTW2} allow to express the twisted wave \eqref{TW} as a function of the cartesian coordinate vector $\bfx$. Starting from the equivalent expression \eqref{TW'} in which the rotation matrix has been diagonalized, we obtain 
\beq
  \bfE_0(\bfx) = \bfN(\bfn) 
  \begin{pmatrix}
     e^{i[(\alpha+1)\varphi+\beta z]} \Jp(\gamma r) \\
     e^{i[(\alpha-1)\varphi+\beta z]} \Jm(\gamma r) \\
     e^{i[\alpha\varphi+\beta z]} J_\alpha(\gamma r) \\
  \end{pmatrix}
  = \bfN(\bfn) \frac{1}{2\pi} \int_0^{2\pi} 
  \begin{pmatrix}
     e^{i(\alpha+1)\Phi} \\ e^{i(\alpha-1)\Phi} \\ e^{i\alpha\Phi} 
  \end{pmatrix}
  e^{i\bfR_\Phi\bfk_0 \cdot \bfx} d\Phi.
\eeq
The geometric meaning of this expression becomes clear by using the following identity which one obtains by applying the intertwining relation \eqref{twine} to the vector $(1,1,1)$: 
\beq \label{twine2}
    \bfN(\bfn) \begin{pmatrix} e^{i\Phi} \\ e^{-i\Phi} \\ 1 \end{pmatrix} = \bfR_\Phi \bfn.
\eeq
This gives 
\beq \label{FourTW3}
   \bfE_0(\bfx) = \frac{1}{2\pi} \int_0^{2\pi} \Bigl( e^{i\alpha\Phi} \bfR_\Phi \bfn\Bigr) 
   e^{i \bfR_\Phi\bfk_0 \cdot \bfx} d\Phi.
\eeq
Expression \eqref{FourTW3} is a Fourier reconstruction integral over a one-dimensional circle in 
$\bfk$-space, 
\beq \label{circle}
     \calC = \{ \bfR_\Phi \bfk_0 \, : \, \Phi\in[0,2\pi]\}
      = \{ \bfk\in\R^3 \, : \, |\bfk| = \frac{\omega}{c}, \; \bfk\cdot \bfe = \beta\},
\eeq
where $\bfe$ is the helical axis. Here we have used that $|\bfk_0|=\omega/c$. See Figure 7. 

\begin{figure} 
\begin{center}
\includegraphics[width=0.4\textwidth]{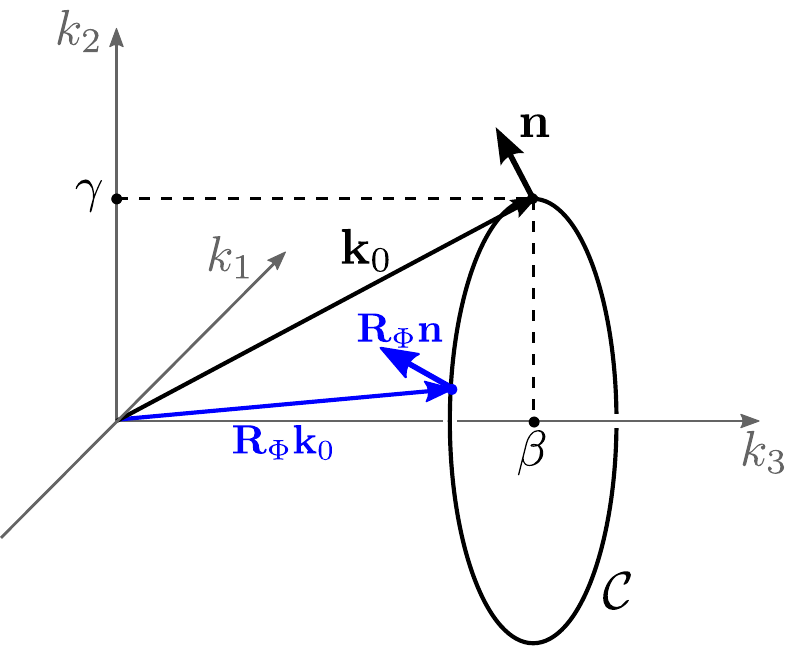}
\caption{Fourier transform of a twisted wave. The circle of wavevectors and associated polarization directions in the picture indicates the plane waves whose phase-modulated superposition gives the twisted wave with parameters $(\alpha,\beta,\gamma)$. The fact that just a {\it one-dimensional} family is needed means that a good approximation can be achieved by a relatively small number of plane waves. This might be of interest for the experimental realization of twisted waves.}
\end{center}
\label{fig:circle}
\end{figure}

Since the circle has radius $\gamma$, the line element (or one-dimensional Hausdorff measure) on $\calC$ is $ds = \gamma \, d\Phi$. Define the following polarization vectorfield as a function of wavevector which corresponds to rotating the polarization along with the base point: 
\beq 
  \tilde{\bfn} \, : \, \calC \to \C^3, \hspace*{1cm} \tilde{\bfn}(\bfR_\Phi \bfk_0) = e^{i\alpha\Phi} \bfR_\Phi \tilde{\bfn}(\bfk_0), \hspace*{1cm} \tilde{\bfn}(\bfk_0)=\bfn.
\eeq 
In terms of this vectorfield on reciprocal space, eq. \eqref{FourTW3} can be written in cartesian reciprocal coordinates as
\beq \label{FourTW4}
   \bfE_0(\bfx) = \frac{1}{(2\pi)^3} \int_{\bfk\in\calC} \frac{(2\pi)^2}{\gamma} \tilde{\bfn}(\bfk)
   e^{i\bfk\cdot \bfx} ds.
\eeq
This is a Fourier reconstruction integral, and hence the Fourier transform of a twisted wave is
\beq \label{FourTW5}
  \widehat{\bfE}_0 (\bfk) = \frac{(2\pi)^2}{\gamma} \tilde{\bfn}(\bfk) \, ds \Big|_{\calC}.
\eeq
Mathematically, this is a singular measure supported on a circle in wavevector space. 

In cylindrical coordinates $(R,\Phi,Z)$ in reciprocal space, the Fourier transform of a twisted wave acquires the simple form
\beq \label{FourTW6}
  \widehat{\bfE}_0(R,\Phi,Z) = \frac{(2\pi)^2}{\gamma} 
  e^{i\alpha\Phi} \bfR_\Phi\bfn \, \delta(R-\gamma)\, \delta(Z-\beta). 
\eeq
Here the radial delta function is, as usual, normalized with respect to $R\, dR$ not $d\, R$, that is to say $\int_0^\infty \delta(R-\gamma) \, R\, dR = 1$. 
\\[2mm]
Formula \eqref{FourTW3} has an interesting alternative interpretation, namely as a {\it group average} instead of a $\bfk$-space integral. Start from the plane wave
\begin{equation} \label{PWFour}
   \bfE_{plane}(\bfx) = \bfn \, e^{i \bfk_0\cdot \bfx},
\end{equation}
and note that the phase and the plane wave in \eqref{FourTW3} satisfy 
$\bfR_\Phi\bfk_0\cdot\bfx = \bfk_0\cdot\bfR_{\Phi}^{-1}\bfx$, $\bfR_\Phi \bfn e^{i\bfk_0\cdot\bfx} = \bfR_\Phi\bfE_{plane}(\bfR_{\Phi}^{-1}\bfx)$. Consequently
\begin{equation} \label{Wig}
   \bfE_0(\bfx) = \int_G \chi_\alpha(g) \Bigl(g\bfE_{plane}\Bigr)(\bfx)\, d\mu(g),
\end{equation}
where $G$ is the group of rotatons around the cylindrical axis, $\chi_\alpha(g)=e^{i\alpha\varphi}$ is a character of the group, and $d\mu=\frac{1}{2\pi}d\Phi$ is the Haar measure on the group, normalized according to the usual convention for compact groups that $\int_G d\mu = 1$. 
This shows that a twisted wave is {\it an integral of the image of a plane wave under the group of rotations about a fixed axis against a character}. The polarization vector $\bfn$ and the ``cartesian reduced wavevector'' $(0,\gamma,\beta)$ are just the polarization vector and the wavevector of this plane wave; the angular wavenumber $\alpha$ comes from the character.

Eq.~\eqref{Wig} is an example of a {\it Wigner projection}, first introduced in the context of quantum systems in \cite{Wigner}. Such a projection maps solutions to 
any linear system of partial differential equations again to solutions provided the system is invariant under the group (as is the case for the time-harmonic Maxwell equations). Moreover it automatically yields solutions to the design equations with group $G$, because for any $h\in G$, formula \eqref{Wig} implies via an elementary change of variables that $h\bfE_0 = \chi_{-\alpha}(h)\bfE_0$. Unfortunately, this simple method to obtain solutions to the design equations cannot be extended in a straightforward manner to non-compact groups which include translations. And projecting any special class of solutions to Maxwell's equations is of course not guaranteed to deliver all solutions to the design equations. 

\section{Conclusions and outlook} \label{sec:outlook}
We have shown on the level of modelling and simulation that twisted X-ray waves would be a very promising tool for structure analysis. 

Numerous theoretical challenges remain. For helical structures, a better understanding of the outgoing radiation in non-axial direction would be desirable. A related issue is to develop a general mathematical theory of the ``radiation transform'', eq. \eqref{rad}, 
which takes the role of the Fourier transform when the incoming radiation is not given by plane waves. Robust phase retrieval algorithms need to be developed for structure reconstruction from twisted X-ray data. And for structures generated by non-abelian symmetry groups such as buckyballs, the right incoming waveforms are not clear; for reasons discussed at the end of Section \ref{sec:design}, in this case the design equations may not be the right approach.

Challenges for the experimental realization include: generation of tunable coherent twisted X-ray waves with a broad spectrum of angular wavenumbers; axial alignment of incoming wave and structure; and achieving a sufficiently strong outgoing signal. 

\section{Appendix: Intensity of outgoing radiation}
Here we relate the intensity \eqref{I} of the real-valued diffracted electromagnetic radiation caused by any incoming time-harmonic field, given by the real part of eq. \eqref{out}, to the {\it complex} electric field amplitude. We rely on the following simple lemma which is applicable due to the special, locally plane-wave-like features that the field vector $\bfE_{out}$ is always perpendicular to $\bfk'$ and $\bfB_{out}$ is proportional to the vector product of $\bfk'$ and $\bfE_{out}$. 
\begin{lemma} {\bf (Poynting vector and intensity of locally plane-wave-like radiation).} 
Suppose that 
$$
    \bfE(\bfx,t)=\mbox{\rm Re}\, (\tilde{\bfE}(\bfx)e^{-i\omega t})), \;\; 
    \bfB(\bfx,t)=\mbox{\rm Re} \, (\bfk(\bfx)\times\tilde{\bfE}(\bfx,t))
$$
for some complex vector field $\,\tilde{\bfE}\, : \, \R^3\to\C^3$ and 
some real vector field $\,\bfk\, : \, \R^3\to\R^3$ with $\bfk(\bfx)\cdot\tilde{\bfE}(\bfx)=0$. Then 
\beq \label{Psimple}
   \bfS = \frac{1}{\mu_0} \bfk \left( 
        |\mbox{\rm Re}\,\tilde{\bfE}|^2 \cos^2 (\omega t) \; + \; |\mbox{\rm Im}\,\tilde{\bfE}|^2\sin^2(\omega t)
     \; + \; 2 \, \mbox{\rm Re}\,\tilde{\bfE}\cdot\mbox{\rm Im}\,\tilde{\bfE} \, \cos(\omega t)\, \sin(\omega t)  
   \right)
\eeq
and
\beq \label{Isimple}
   I = \frac{1}{2\mu_0}\,  |\bfk|\,  |\tilde{\bfE}|^2.
\eeq
\end{lemma} 
Thus the Poynting vector of the real part of a locally plane-wave-like electromagnetic field is a
fluctuating mixture of contributions from both the real and imaginary parts of the spatial electric field, and the
intensity is proportional to the absolute value squared of the {\it complex} electric field amplitude. One 
can hence say that the mystery of complex field amplitudes appearing in intensities of real, physical fields is
explained by time-averaging.  
\\[2mm]
{\bf Proof} Writing $\mbox{Re}\,\tilde{\bfE}=\bfE_1$, $\mbox{Im}\,\tilde{\bfE}=\bfE_2$, we have
$\bfE = \bfE_1\cos(\omega t) - \bfE_2\sin(\omega t)$, 
$\bfB = (\bfk\times\bfE_1)\cos(\omega t) - (\bfk\times\bfE_2)\sin(\omega t)$.
Eq. \eqref{Psimple} now follows from the expansion rule for double vector products, 
$\bfa\times(\bfb\times\bfc) = \bfb(\bfa\cdot\bfc) - \bfc(\bfa\cdot\bfb)$, and the orthogonality assumption 
$\bfk\cdot\bfE_1=\bfk\cdot\bfE_2=0$. 
Moreover since the time-dependent factor in \eqref{Psimple} is always nonnegative, the absolute value $|\bfS|$ is
a scalar multiple of this factor and it is straightforward to evaluate the asymptotic time average 
in \eqref{I}, which amounts to averaging over one period. Since $\cos^2$ and $\sin^2$ have average $1/2$
and $\cos\cdot \sin$ has average $0$, we obtain \eqref{Isimple}.
\\[2mm]
The intensity of diffracted physical radiation at the observation point $\bfx$, i.e. the intensity of the
real part of the elecromagnetic field \eqref{out}, is therefore, by eq. \eqref{Isimple}, 
$$
        I(\bfx) = \frac{c\,\eps_0}{2} \, |\bfE_{out}(\bfx,t)|^2 \;\;\; (\mbox{eq. \eqref{Iout}, Section \ref{model}}),
$$
where $\bfE_{out}$ is the complex electric field in \eqref{out} (note that its absolute value is independent of $t$). 
Here we have used the relations $|\bfk'|=\omega/c$ and $\eps_0\mu_0=1/c^2$.

\vspace{5mm}
\noindent {\bf Acknowledgments}. GF was partially supported by DFG through SFB-TR 109. RDJ was supported by AFOSR (FA9550-15-1-0207) and partially supported by ONR (N00014-14-0714), NSF/PIRE (OISE-0967140), the MURI program (FA9550-12-1-0458), and
a John Von Neumann visiting professorship at TU Munich. DJ was partially supported by a stipend from Universit\"at Bayern e.V. 
$\,$GF thanks Nick Schryvers for helpful discussions on vortex beams, John Ockendon for helpful discussions which led to the spectral interpretation of the twisted wave parameters, and Robin Santra for providing advice and reference \cite{Santra} on the derivation of X-ray diffraction intensities from quantum electrodynamics.      

\begin{small}

\end{small}


\begin{thebibliography}{99}

\bibitem[AB92]{AB92} {\sc L. Allen and M. W. Beijersbergen and R. J. C. Spreeuw and J. P. Woerdman}, {\em Orbital angular momentum of light and the transformation of Laguerre-Gaussian laser modes}, Phys. Rev. Lett. 45 (1992), 8185--8189.

\bibitem[AD00]{AD00} {\sc J. Arlt and K. Dholakia}, {\em Generation of high-order Bessel beams by use of an axicon}, Optics Communications 177 (2000), 297--301.

\bibitem[AM11]{Nielsen} {\sc J. Als-Nielsen and D. McMorrow}, {\em Elements of Modern X-ray Physics}, John Wiley and Sons, New York (2011).

\bibitem[AM76]{Ashcroft} {\sc N.~W. Ashcroft and N.~D. Mermin}, {\em Solid State Physics}, Saunders College Publishing (1976).

\bibitem[BG08]{BaakeGrimm} {\sc M. Baake and U. Grimm}, {\em The singular continuous diffraction measure of the Thue–-Morse chain}, J. Phys. A: Math. Theor. 41 (2008), 422001

\bibitem[BG13]{BaakeGrimmBook} {\sc M. Baake and U. Grimm}, {\em Aperiodic Order Vol. 1: A Mathematical Invitation}, Cambridge University Press (2013).

\bibitem[BM04]{BaakeMoody} {\sc M.~Baake and R.~Moody}, {\em Weighted Dirac combs with pure point diffraction}, J. Reine und Angew. Math. (Crelle) 573 (2004), 61--94.

%\bibitem[BCX12]{Ball} {\sc J.~M. Ball, Y. Capdeboscq, B.~T. Xiao}, {\em On uniqueness for time harmonic anisotropic Maxwell's equations with piecewise regular coefficients}, Math. Models Methods Appl. Sci 22 (2012), 1250036.

%\bibitem[Br13]{Bragg} {\sc W.~L. Bragg}, {\em The Structure of Some Crystals as Indicated by Their Diffraction of X-rays}, Proc. R. Soc. Lond. A 89 (1913), 248--277.

\bibitem[CCV52]{CCV52} {\sc C. and F.~H.~C. Crick and V.~Vand}, {\em The Structure of Synthetic Polypeptides. I. The Transform of Atoms on a Helix}, Acta Cryst. 5 (1952), pp. ~581--586.

\bibitem[CK62]{Caspar} {\sc D. L.~D. Caspar and A. Klug}, {\em Physical principles in the construction of regular viruses}, Cold Spring Harbor Symp. Quant. Biol. {\bf 27} (1962), 1--24.

%\bibitem[Ch06]{Chapman} {\sc H.~N. Chapman, et al.}, {\em Femtosecond diffractive imaging with a soft-X-ray free-electron laser}, Nature Physics 2 (2006), 839--843.

%\bibitem[Co95]{Cowley} {\sc J.~M. Cowley}, {\em Diffraction Physics}, North-Holland Personal Library (1995).

\bibitem[CT97]{CT} {\sc C. Cohen-Tannoudji and J. Dupont-Roc and G. Grynberg}, 
{\em Photons and Atoms. Introduction to Quantum Electrodynamics}, Wiley (1997)

\bibitem[DEJ]{JamesDayal} {\sc K. Dayal, R. Elliott, and R.~D. James}, {\em Objective Formulas}, unpublished.

\bibitem[Di75]{Di75} {\sc J. Dieudonn\'e}, {\em \'El\'ements d'analyse}, vol. 6, Gauthier-Villars, Paris (1975).

\bibitem[FJJ15]{FJJ2} {\sc G. Friesecke, R.~D. James, and D. J\"ustel}, in preparation.

\bibitem[FKL12]{FKL12} {\sc W. Friedrich, P. Knipping, and M. von Laue}. {\em Interferenzerscheinungen bei R\"ontgenstrahlen}, Sitzungsberichte der Mathematisch-Physikalischen Classe der K\"oniglich-Bayerischen Akademie der Wissenschaften zu M\"unchen, 1912.

\bibitem[Fr07]{Friesecke} {\sc G. Friesecke}, {\em Lectures on Fourier Analysis}, lecture notes, University of Warwick (2007).

\bibitem[GZ11]{Ge} {\sc P. Ge, Z. H. Zhou}, {\em Hydrogen-bonding networks and RNA bases revealed by cryo electron microscopy suggest a triggering mechanism for calcium switches}, Proc. Natl. Acad. Sci. {\bf 108} (2011), 9637–-9642.

%\bibitem[Gu63]{Guinier} {\sc A. Guinier}, {\em X-Ray Diffraction}, W.~H. Freeman and Company (1963).

\bibitem[Gr99]{Griffiths} {\sc D.~J. Griffiths}, {\em Electromagnetic Theory}, Prentice Hall (1999).

\bibitem[Ha34]{Hansen}
{\sc W.~W. Hansen}, {\em Transformations Useful in Certain Antenna Calculations}, J. Appl. Phys. 8 (1937), 282--286.

\bibitem[HD92]{HD92} {\sc N.~R. Heckenberg and R.~McDuff and C.~P. Smith and A.~G. White}, {\em Generation of optical phase singularities by computer generated holograms}, Opt. Lett. 17 (1992), 221--223.

\bibitem[HR63]{HewittRoss} {\sc E. Hewitt and K.~A. Ross}, {\em Abstract Harmonic Analysis I}, Springer (1963). 

\bibitem[Ho95]{Hof}
{\sc A. Hof}, {\em On diffraction by aperiodic structures}, Comm. Math. Phys. 169 (1995), 25--43. 

\bibitem[Ja98]{Jackson} {\sc J.~D. Jackson}, {\em Classical Electrodynamics}, John Wiley and Sons, New York (1998).

\bibitem[Ja06]{James} {\sc R.~D. James}, 
{\em Objective structures}, J. Mech. Phys. Solids 54 (2006), 2354--2390. 

\bibitem[JJF15]{FJJ3} {\sc D. J\"ustel, R.~D. James, and G. Friesecke}, in preparation.

\bibitem[Ju15]{Juestel} {\sc D. J\"ustel}, in prepration.

\bibitem[KCW58]{KlugEtAl} {\sc A. Klug and F.~H.~C. Crick and H.~W. Wyckhoff}, {\em Diffraction by Helical Structures}, Acta Cryst. 11 (1958), 199--213.

\bibitem[La15]{Lasser} {\sc R. Lasser}, Monograph on hypergroups, in preparation.
  
%\bibitem[MCKS99]{MCKS99} {\sc J. Miao, P. Charalambous, J. Kirz, and D. Sayre}, {\em Extending the methodology of X-ray crystallography to allow imaging of micrometer-sized non-crystalline specimens}, Nature 400 (1999), 342--344.

%\bibitem[MSS12]{MSS12} {\sc J. Miao, R.~L. Sandberg, and C. Song}, {\em Coherent X-Ray Diffraction Imaging}, IEEE J. Sel. Top. Quant., 18 (2012), pp.~399--410.

\bibitem[MTT07]{MTT07} {\sc G. Molina-Terriza and J.~P. Torres and L Torner}, 
{\em Twisted photons}, Nature Physics 3, (2007), 305--310
  
\bibitem[RS00]{Reiter} {\sc H. Reiter, and J.~D. Stegeman}, {\em Classical Harmonic Analysis and Locally Compact Groups}, Oxford University Press (2000).

%\bibitem[Sa52a]{Say52a}{\sc D. Sayre}, {\em The squaring method: a new method for phase determination}, Acta Cryst., 5 (1952), 60--65.

%\bibitem[Sa52b]{Say52b} {\sc D. Sayre}, {\em Some implications of a theorem due to Shannon}, Acta Cryst., 5 (1952), 843.

\bibitem[Sa09]{Santra} {\sc R. Santra}, {\em Concepts in X-ray physics}, J. Phys. B: At. Mol. Opt. Phys. 42 (2009), 023001. 

%\bibitem[SBGC84]{SBGC84} {\sc D. Shechtman, I. Blech, D. Gratias, J. Cahn}, {\em Metallic Phase with Long-Range Orientational Order and No Translational Symmetry}, Phys. Rev. Lett. 53 (1984), 1951--1954.

\bibitem[St94]{Strichartz} {\sc R. Strichartz}, {\em A Guide to Distribution Theory and Fourier Transforms}, CRC Press, (1994).

%\bibitem[St01]{Stubbs} {\sc G. Stubbs}, {\em Fibre diffraction studies of filamentous viruses}, Rep. Prog. Phys. {\bf 64} (2001), 1389--1425.

\bibitem[UT10]{UT10} {\sc M. Uchida and A. Tonomura}, 
{\em Generation of Electron Beams Carrying Orbital Angular Momentum}, Nature 464 (2010), 737--739.

\bibitem[VTS10]{VTS10} {\sc J. Verbeeck, H. Tian, and P. Schattschneider}, {\em Production and application of electron vortex beams}, Nature 467 (2010), pp.~301--304.

\bibitem[We64]{Weil} {\sc A. Weil}, {\em Sur certains groupes d'op\'erateurs unitaires}, Acta Math. 111 (1964), pp.~143-211.

\bibitem[Wi31]{Wigner} {\sc E. Wigner}, {\em Gruppentheorie und ihre Anwendung auf die Quantenmechanik der Atomspektren}, Vieweg (1931).

\end{thebibliography}
\end{document}